\documentclass[aps,pra,reprint,superscriptaddress,longbibliography]{revtex4-1}

\usepackage{hyperref}
\usepackage{graphicx}  
\usepackage{bm}        
\usepackage{amssymb}   
\usepackage{xspace}
\usepackage{verbatim}
\usepackage{color}
\usepackage{soul}
\usepackage{subfigure}
\usepackage[export]{adjustbox}
\usepackage{dcolumn}
\usepackage{amsthm,stmaryrd,mathtools}
\usepackage{txfonts}
\usepackage{braket}
\usepackage{amsmath}
\usepackage{xcolor}
\usepackage{array}
\usepackage{multirow}
\usepackage{tabularx}
\usepackage{booktabs} 
\usepackage{lipsum}
\usepackage{dsfont}
\definecolor{lightblue}{RGB}{100,100,250}

\begin{document}

\global\long\def\ket#1{\left|#1\right\rangle }%
\global\long\def\bra#1{\left\langle #1\right|}%
\global\long\def\braket#1#2{\langle#1|#2\rangle}%
\global\long\def\expectation#1#2#3{\langle#1|#2|#3\rangle}%
\global\long\def\average#1{\langle#1\rangle}%

\author{Rozhin Yousefjani}%
\email{ryousefjani@hbku.edu.qa}
\affiliation{Qatar Center for Quantum Computing, College of Science and Engineering, Hamad Bin Khalifa University, Doha, Qatar}

\author{Saif Al-Kuwari}%
\affiliation{Qatar Center for Quantum Computing, College of Science and Engineering, Hamad Bin Khalifa University, Doha, Qatar}

\title{Mechanical Squeezed-Fock Gravimeter}

\begin{abstract}
Levitated mechanical systems are promising candidates for quantum gravimetry, as gravity couples directly to their center-of-mass motion, enabling the large mass of a mesoscopic particle to serve as a sensing resource. In this paper, we propose a mechanical squeezed-Fock qubit gravimeter using a Duffing oscillator that is driven by a detuned two-phonon pump. In the squeezed-Fock basis, the gravitational force 
couples to the anti-squeezed quadrature, which enhances the gravity-induced transition rate while preserving the direct mass scaling of the mechanical force coupling. We show that sensitivity improves with reduced effective qubit splitting that is controlled by the squeezing parameter and the Duffing nonlinearity. We further analyze mechanical damping and show that squeezing converts ordinary dissipation into anisotropic qubit noise, setting a practical trade-off between signal amplification and decoherence rate. These results identify the mechanical squeezed-Fock qubit as a new platform for quantum-enhanced gravimetry.
\end{abstract}

\maketitle

\section{Introduction}

Precision measurements of gravitational acceleration $g$ lie at the intersection
of fundamental physics and applied science, underpinning applications that range
from underground mapping and geophysical surveying~\cite{Stray2022QuantumGravityCartography,crawford2025quantum,forster2021geothermal,Wu2019GravitySurveys} to inertial
navigation~\cite{krelina2021quantum} and tests of the equivalence principle~\cite{ye2024essay,poli2011precision}. Generally, gravimeters fall
into two distinct classes: \emph{absolute} gravimeters, which determine
$g$ at a single point independent of any other measurement, and \emph{relative}
gravimeters, which register only a change or difference in $g$, between two
locations, two times, or two states of the same apparatus. The leading absolute
technology is the free-falling cold-atom interferometer, which encodes the
gravitational phase into the coherence of atomic matter waves~\cite{kasevich1991atomic,peters1999measurement,Geiger2020HighAccuracyColdAtomSensors,wei2025adaptive}, reaching
sub-$\mu$Gal/$\sqrt{\rm Hz}$ ($1\, \mu$Gal${=} 10\,$nm/s$^2$) sensitivities~\cite{bidel2018absolute,Hu2013UltrahighSensitivityGravimeter,Menoret2018TransportableAQG,Zhang2023BraggAtomGravimeter}. However, this sensitivity
scales with the interrogation arm length and requires
free-fall operation~\cite{Peters2001HighPrecisionGravity,Geiger2020HighAccuracyColdAtomSensors,LeGouet2008LimitsCompactAtomicGravimeter,Merlet2009OperatingBeyondLinearRange,Zhou2012ActiveVibrationIsolator,Menoret2018TransportableAQG,Geiger2020HighAccuracyColdAtomSensors}.
\\ \\
The majority of gravimetric applications, however, do not require an absolute
value of $g$ at every measurement point. Monitoring volcanic and seismic
activity, tracking aquifer and hydrocarbon reservoirs, and dense-network
geodetic surveying instead require detecting small \emph{changes} in $g$ with
high sensitivity, and, increasingly, reduced size.
This is the historical role of relative gravimeters: spring-based mechanical
gravimeters and superconducting-sphere gravimeters, in which a levitated
superconducting~\cite{griggs2017sensitive,liu2025progress} test mass tracks small deviations in $g$, are
calibrated periodically against absolute stations~\cite{rademacher2020quantum}.
\\ \\
Levitated mesoscopic particles have emerged as a compelling platform for this
second class of instrument~\cite{rademacher2020quantum,gonzalez2021levitodynamics,bose2025massive}. Unlike a free-falling atom, a levitated
particle is held in a trap whose restoring potential must, at equilibrium,
balance the gravitational force acting on it; the particle's response to $g$ is
therefore encoded as a small perturbation around this compensated operating
point. Levitated-particle sensors are
thus naturally suited to the same measurement task as spring and superconducting
gravimeters, but with two decisive advantages over both: the absence of
mechanical clamping enables quality factors as high as $Q{\sim}10^9$ under high
vacuum~\cite{dania2024ultrahigh}, and the gravitational response of the levitated object is
proportional to its own mass $m$, giving a susceptibility that scales as
$\sqrt{m}$ orders of magnitude larger than for atomic sensors~\cite{rademacher2020quantum,Huo2026QuantumGravimetryMechanicalQubits}. 
The most natural approach is to encode $g$ directly in the center-of-mass (CM) motion and read out the resulting phase or population~\cite{delic2020cooling,tebbenjohanns2021quantum,magrini2021realtime,kamba2025squeezing,Marti2024QuantumSqueezing,dania2024ultrahigh,piotrowski2023simultaneous}. 
However, the CM mode of a harmonic oscillator has an equally spaced spectrum with no metrological advantage specific to a particular pair of levels. To solve this issue, conventional levitated-particle gravimetry has therefore coupled the mechanical mode to other strongly nonlinear quantum systems, such as an electron-spin qubit~\cite{scala2013matterwave,wang2025enhanced} or an optical cavity~\cite{qvarfort2018gravimetry}.
However, this approach has fundamental limitations.
The coupling between the CM mode and the auxiliary system scales as $1/\sqrt{m}$~\cite{li2020enhancing,li2016hybrid}, exactly canceling the $\sqrt{m}$ enhancement from the particle's mass. As a result, the sensitivity of hybrid gravimeters is mass-independent~\cite{scala2013matterwave,wang2025enhanced,qvarfort2018gravimetry}.
\\ \\ 
A possible solution is to encode the qubit directly in the CM mode, established based on quantized motional modes, provided that the mechanical spectrum is made sufficiently anharmonic~\cite{Rips2013NanomechanicalQubits}. 
Introducing Duffing nonlinearity splits the ladder of Fock states and allows the two lowest levels to be addressed selectively~\cite{Pistolesi2021NanomechanicalQubit,Samanta2023NonlinearNanomechanical,Sharma2025Towards}. This \emph{mechanical qubit} (MQ)~\cite{Savelev2006BucklingNanobars,Savelev2007QuantumTunnelingNanobars,Rips2013NanomechanicalQubits,Rips2014NonlinearNanomechanical,Pistolesi2021NanomechanicalQubit,Fluehmann2019TrappedIonMechanicalOscillatorQubit} has recently been demonstrated experimentally~\cite{Yang2024MechanicalQubit}, establishing phononic two-level systems as a realistic platform for quantum information~\cite{Pistolesi2024Journey} and quantum sensing~\cite{Pistolesi2021NanomechanicalQubit,Qiao2026MechanicalSqueezedFockQubit,Huo2026QuantumGravimetryMechanicalQubits}. 
Since the MQ couples to gravity directly through its CM displacement, with no auxiliary system involved, the mass enhancement is fully preserved. Huo \emph{et al.}~\cite{Huo2026QuantumGravimetryMechanicalQubits} recently exploited this insight to propose a mass-enhanced quantum gravimeter based on MQs and their cat-state counterparts.
Despite this progress, MQ implementations remain constrained by a central physical bottleneck: nanomechanical resonators typically exhibit weak intrinsic nonlinearities and limited anharmonicity, making robust state preparation and coherent control challenging~\cite{Rips2013NanomechanicalQubits,Pistolesi2021NanomechanicalQubit,Samanta2023NonlinearNanomechanical,Yang2024MechanicalQubit,Marti2024QuantumSqueezing}.
A complementary line of development is to map the original Fock ladder into a \emph{squeezed-Fock ladder} whose effective anharmonicity is exponentially enhanced by $e^{4r}$, where $r$ is the squeezing parameter~\cite{Qiao2026MechanicalSqueezedFockQubit}. The resulting \emph{mechanical squeezed-Fock qubit} (MSFQ) allows a mechanical qubit to be encoded even when the intrinsic nonlinearity is weak~\cite{Qiao2026MechanicalSqueezedFockQubit}. 
\\ \\ 
In this paper, we propose and analyze the mechanical squeezed-Fock gravimeter, a new sensing approach that combines the direct mass advantage of the MQ architecture with the exponential signal amplification of the squeezed-Fock states. We drive the CM mode with a \emph{detuned} two-phonon pump and employ a Rabi-based interrogation protocol sensitive to the static gravity force. Gravity couples to the anti-squeezed quadrature of the CM mode with an amplitude ${\propto}e^r$, producing a gravity-induced coupling that exceeds its MQ counterpart~\cite{Huo2026QuantumGravimetryMechanicalQubits} by $e^r$. In our protocol, effective qubit splitting can be fully controlled by adjusting the Duffing nonlinearity as well as the squeezing parameter, which extends the coherent interrogation time and further improving sensitivity. Using quantum estimation theory~\cite{Helstrom1969QuantumDetection,Paris2009QuantumEstimation,Montenegro2025Review}, we derive the exact quantum Fisher information (QFI) and show that it is saturated by a simple squeezed-Fock population measurement. 
We further analyze the effect of mechanical damping, caused by both zero- and finite-temperature environments, and show that squeezing transforms ordinary damping into anisotropic qubit noise in the squeezed-qubit subspace.
As a result, the standard population measurement is no longer optimal, and the optimized readout is obtained by measuring along the symmetric-logarithmic-derivative direction.
The performance of the gravimeter is governed by a competition between 
coherent Rabi precession and the anisotropic decay channel.
We determine the practical operating window of the sensor and show that squeezing is beneficial as long as the enhanced decoherence rate remains comparable to the coherent qubit dynamics.

\section{Quantum estimation theory}
\label{sec:quantum_metrology}

For an unknown parameter $g$ that is encoded in the sensor state $\rho_g(t)$, a measurement, represented by a POVM $\{\Pi_x\}$, gives the outcome probabilities as $p(x|g){=}\mathrm{Tr}[\rho_g(t)\Pi_x]$. For this fixed measurement, the classical Fisher information (CFI) is $\mathcal F_{\rm C}(g){=}\sum_x [\partial_g p(x|g)]^2/p(x|g)$, with zero-probability outcomes omitted. For $\nu$ independent repetitions, any unbiased estimator satisfies $\delta g{\geq} 1/\sqrt{\nu\mathcal F_{\rm C}(g)}$. Optimizing over all POVMs gives the QFI, $\mathcal F_Q(g){=}\max_{\{\Pi_x\}}\mathcal F_{\rm C}(g)$.
The corresponding quantum Cram\'er--Rao bound is $\delta g\geq 1/\sqrt{\nu\mathcal F_Q(g)}$. 
The QFI can be written in terms of the SLD $\hat L_g$, defined implicitly by $\partial_g\rho_g{=}(\rho_g\hat L_g{+}\hat L_g\rho_g)/2$. It is then $\mathcal F_Q(g){=}\mathrm{Tr}[\rho_g\hat L_g^2]
{=}\mathrm{Tr}[(\partial_g\rho_g)\hat L_g].$
The locally optimal measurement is the projective measurement in the eigenbasis of $\hat L_g$.
For a pure state $\rho_g{=}|\psi_g\rangle\langle\psi_g|$, the QFI reduces to~\cite{braunstein1994statistical}
\begin{equation}
\mathcal F_Q(g)
=4\left[
\langle\partial_g\psi_g|\partial_g\psi_g\rangle
-\left|\langle\psi_g|\partial_g\psi_g\rangle\right|^2
\right].
\label{eq:QFI_pure_main}
\end{equation}
For a qubit mixed state $\rho_g{=}(\mathbb I{+}\bm v\cdot\bm\sigma)/2$, one has $\mathcal F_Q(g){=}|\partial_g\bm v|^2+(\bm v\cdot\partial_g\bm v)^2/(1-|\bm v|^2)$ for $|\bm v|<1$.

\section{Model}

The MQ considered here is made from the CM motion of a levitated mesoscopic particle along the gravitational direction with Duffing nonlinearity of strength $D$. The Duffing nonlinearities are naturally present in levitated optomechanical systems and have been experimentally characterized~\cite{Yang2024MechanicalQubit}. The free Hamiltonian of this mechanical mode is written as
\begin{equation}
\hat H
 {=} 
\hbar \omega \hat a^\dagger \hat a
-
\hbar D \hat a^{\dagger} \hat a^{\dagger}\hat a \hat a,
\label{eq:Hm_append}
\end{equation}
where $\hat a$ ($\hat a^\dagger$) is the annihilation (creation) operator of the CM mode with frequency $\omega$. The corresponding eigenenergies of this system are
$E_n {{=}} n\hbar \omega - n(n-1)\hbar D.$
Clearly, unlike a purely harmonic oscillator, the level spacing is no longer uniform. In particular, the transition frequency between the Fock states $|1\rangle$ and $|2\rangle$ differs from that between $|0\rangle$ and $|1\rangle$ by the anharmonicity $ {-}2\hbar D$.
This anharmonicity is the key ingredient that allows one to interpret the mechanical oscillator as an effective qubit rather than as a fully harmonic mode.
A MQ is obtained when the system dynamics is effectively restricted to the two-dimensional subspace spanned by the lowest two eigenstates, namely the ground state $|0\rangle$ and the first excited state $|1\rangle$. Physically, this requires that all external perturbations, driving amplitudes, and dissipative broadenings remain much smaller than the anharmonicity value, so that leakage to higher states is suppressed. 
\\ \\
To create a squeezed-Fock spectrum following~\cite{Qiao2026MechanicalSqueezedFockQubit}, we drive our Duffing oscillator with a \emph{detuned} two-phonon pump of frequency $2\omega_p$ with phase $\theta$ and amplitude $A_p$. This reshapes the CM spectrum into a squeezed and more strongly anharmonic one, such that the effective qubit subspace is no longer formed by the bare Fock states $\{|0\rangle, |1\rangle\}$, but by the two lowest squeezed-Fock states.  
The Hamiltonian of the Duffing oscillator under detuned drive reads
\begin{align}
\hat H
 &=
\hbar \omega \hat a^\dagger \hat a
-
\hbar D \hat a^{\dagger}\hat a^{\dagger}\hat a \hat a \nonumber\\
&+
\frac{\hbar A_p}{2}
\left(
e^{-i(2\omega_p t+\theta)} \hat a^2
+
e^{i(2\omega_p t+\theta)} \hat a^{\dagger 2}
\right). 
\label{eq:Hlab}
\end{align} 
In the frame rotating at $\omega_p$ via $\hat U_0(t){{=}}e^{-i\omega_p \hat a^\dagger \hat a t}$, one has
\begin{align}\label{eq:Hrot} 
\hat H_{\rm rot}
= 
\hbar \delta \, \hat a^\dagger \hat a
-
\hbar D \hat a^{\dagger}\hat a^{\dagger}\hat a \hat a 
+ 
\frac{\hbar A_p}{2}
\left(
e^{-i\theta}\hat a^2
+
e^{i\theta}\hat a^{\dagger 2}
\right)
\end{align}
where $\delta{{=}}\omega{-}\omega_p $ is the detuning.
Using the Bogoliubov transformation $\hat b {{=}} \hat S \hat a \hat S^{\dagger}$ with the squeezing operator $\hat S {{=}} \exp(r^* \hat a^2/2 {+} r \hat{a}^{\dagger 2}/2 )$ and squeezing parameter $r$ that satisfies $\tanh(2r){{=}}A_p/\delta$
one can obtain the new mechanical mode $\hat b {{=}} \cosh r \, \hat a + e^{i\theta}\sinh r \, \hat a^\dagger$.
Equivalently one has 
$\hat a {{=}} \cosh r \, \hat b - e^{i\theta}\sinh r \, \hat b^\dagger$.
Under the rotating-wave approximation (RWA), the Hamiltonian~(\ref{eq:Hrot}) is transformed into an effective form as (see Appendix~\ref{app:Effective Hamiltonian}),
\begin{align}
\hat H_{\rm eff}
&\simeq
\hbar \omega_b \hat b^\dagger \hat b
-\hbar U_b \hat b^{\dagger }\hat b^{\dagger }\hat b\hat b,
\label{eq:HG_RWA}
\end{align}
with $\omega_b
{{=}} \sqrt{\delta^2-A_p^2} {-} D\!\left( 8\cosh^2 r \sinh^2 r + 4\sinh^4 r \right),$
and $U_b {{=}} \frac{D}{4}\left[3\cosh(4r)+1\right].$
The minus sign in front of the quartic term reflects the sign of the Duffing nonlinearity in the levitated mechanical platform. The shape of $\omega_b$ obligates us to determine the boundary at which the qubit gap closes, namely $\omega_b{=}0$. As such, we introduce $D_{\rm crit}{=}\sqrt{\delta^2-A_p^2} \Big/ \left( 8\cosh^2 r \sinh^2 r + 4\sinh^4 r \right)$ and define the physically allowed range of the Duffing parameter $D{<}D_{\rm crit}$ for which $\omega_b$ does not vanish. Besides this, the RWA validity limits the accessible range of $D/D_{\rm crit}$. Beyond this region, the effective Hamiltonian~\eqref{eq:HG_RWA} is no longer a reliable description of the dynamics as the extra terms of the Hamiltonian, corresponding to rapid oscillations, become important, causing additional frequency shifts and leakage out of the squeezed-qubit subspace, so the effective Hamiltonian in Eq.~(\ref{eq:HG_RWA}) is no longer quantitatively reliable.
See Appendix~\ref{app:Effective Hamiltonian} for a detailed discussion.

\section{Quantum Gravimeter}
In the mechanical-mode description, gravity appears as a static linear force,
\((mg{-}F)x_0(\hat a+\hat a^\dagger)\). 
The detuned two-phonon pump defines a squeezed-Fock basis in which the displacement quadrature coupled to gravity is rescaled by \(e^r\), provided that the pump phase $\theta$ is chosen such that gravity couples to the anti-squeezed quadrature. 
Selecting $\theta{{=}}\pi,$ results in $\hat a{+}\hat a^\dagger {{=}} e^r(\hat b{+}\hat b^\dagger)$, and hence, $\hat H_g^{\rm s} {{=}} G (\hat b{+}\hat b^\dagger)$, with $G{{=}}e^r(mg{-}F) x_0$. Here, $F$ is a static compensation force applied opposite to gravity and chosen such that $G{\ll}2U_b$, which keeps the sensor near its optimal working point and suppresses leakage out of the squeezed-qubit subspace. 
The protocol measures deviations in $g$ relative to the fixed offset defined by $F$.
The quantity $x_0{{=}}\sqrt{\hbar / 2m\omega}$ is the zero-point fluctuation amplitude.
In the MSFQ framework, where $\hat b+\hat b^\dagger{\rightarrow} \hat \sigma_x$, the effective gravimeter Hamiltonian reads
\begin{equation}
\hat H_{\rm eff}^{\rm s}
{{=}}
\frac{\hbar \omega_b}{2}\hat \sigma_z
+
\frac{\hbar \Omega_g^{\rm s}}{2} \hat \sigma_x,
\label{eq:HeffS}
\end{equation}
with the gravity-induced coupling $\Omega_g^{\rm s} {{=}} 2e^r \frac{(mg-F)x_0}{\hbar}.$
Obviously, the static gravitational force is enhanced by a factor $e^r$ at the Hamiltonian level, which makes the qubit populations more sensitive to small changes in $g$. This enhancement does not come from adding an auxiliary sensor or increasing the particle mass. It comes from reshaping the mechanical mode itself.
\\ \\
Initializing the sensor in the ground squeezed state, $|\psi(0)\rangle {{=}} |0\rangle_{\rm s},$ results in $|\psi(t)\rangle
 {=} e^{-i \hat H_{\rm eff}^{\rm s}t/\hbar} \, |0\rangle_{\rm s}
 {=}  \alpha(t) |0\rangle_{\rm s} + \beta(t)|1\rangle_{\rm s},$
where $\alpha(t){=}
\cos\left(\frac{\omega_{\rm R}t}{2}\right)
+
i\frac{\omega_b}{\omega_{\rm R}}
\sin\left(\frac{\omega_{\rm R}t}{2}\right),$ and $\beta(t){=}
-i\frac{\Omega_g^{\rm s}}{\omega_{\rm R}}
\sin\left(\frac{\omega_{\rm R}t}{2}\right)$ for $\omega_{\rm R}
{{=}}
\sqrt{\omega_b^2+\Omega_g^{\rm s2}}$
as the generalized Rabi frequency. 
The exact QFI for estimating $g$ from this state is derived in the Appendix~\ref{app:coherent_fisher}. In the weak-force regime, namely $\Omega_g^{\rm s}\ll \omega_b$, the exact result reduces to
\begin{equation}
\mathcal{F}_Q(g,t)
\simeq
\frac{8me^{2r}}{\hbar\omega\omega_b^2}
\sin^2\left(\frac{\omega_b t}{2}\right).
\label{eq:FQ_g_weak_x0_main}
\end{equation}
which peaks at $\omega_b t{=}(2\ell+1)\pi$, with integer $\ell$. The first optimal interrogation time is $t_{\rm opt} = \frac{\pi}{\omega_b}$
for which
\begin{equation}
\mathcal{F}_{Q,{\rm max}}(g)
\simeq
\frac{8me^{2r}}{\hbar\omega\omega_b^2}.
\label{eq:FQ_g_max_weak_main}
\end{equation}
If each run takes an interrogation time $t$ and the total averaging time is $T$, then $\nu{=}T/t$ and the time-normalized sensitivity obeys
\begin{equation}
\sqrt{T}\delta g^{\rm s}_{\rm opt}
\simeq
\sqrt{
\frac{\pi\hbar\omega\omega_b}
{8m e^{2r}}
}
\label{eq:QCRBS_opt_weak_main}
\end{equation}
at $t{=}t_{\rm opt}$.
To approach this bound, one can measure the excited squeezed-Fock population, $\hat O_{\rm s}{=}|1\rangle_{\rm s}{}_{\rm s}\langle 1|.$
The corresponding probability is $P_1(t) {=}\frac{\Omega_g^{\rm s2}}{\omega_{\rm R}^2}
\sin^2\left(\frac{\omega_{\rm R}t}{2}\right).$
The exact CFI associated with the binary outcomes $P_1(t)$ and $1{-}P_1(t)$ is given in the Appendix~\ref{app:coherent_fisher}. In the same weak-force regime, it reduces to
\begin{equation}
\mathcal{F}_{\rm C}(g,t)
\simeq
\frac{8me^{2r}}{\hbar\omega\omega_b^2}
\sin^2\left(\frac{\omega_b t}{2}\right)
=
\mathcal{F}_Q(g,t).
\label{eq:CFI_weak_main}
\end{equation}
Therefore, the excited-state occupation measurement is optimal in the weak-force regime. 
\\ 
Apart from $m$ and $\omega$, the precision is governed most directly by two effective quantities: the squeezing parameter $r$ and the effective qubit frequency $\omega_b$. While $r$ is only controlled by the pump ratio $A_{p}/\delta$, the frequency $\omega_b$ is jointly controlled by $\delta$, $A_{p}/\delta$, and $D/D_{\rm crit}$. 
Obviously, the sensitivity is improved by increasing the Hamiltonian-level signal enhancement $e^r$ and by reducing $\omega_b$, but only within a regime where the interrogation time remains finite, and the RWA is valid, see Appendix~\ref{app:Effective Hamiltonian}.
\begin{figure}[t!]
    \centering
    \includegraphics[width=0.49\linewidth]{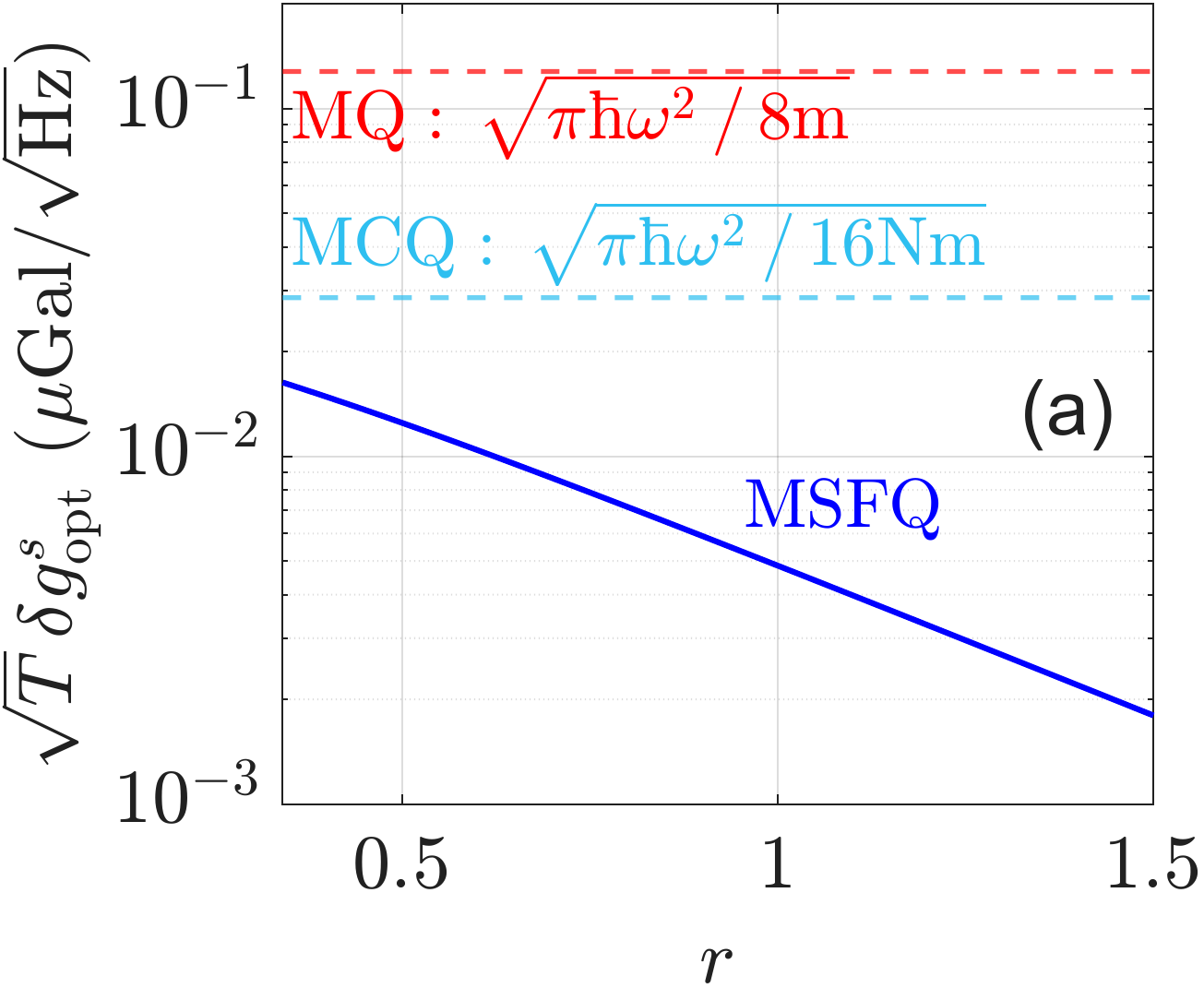}
    \includegraphics[width=0.49\linewidth]{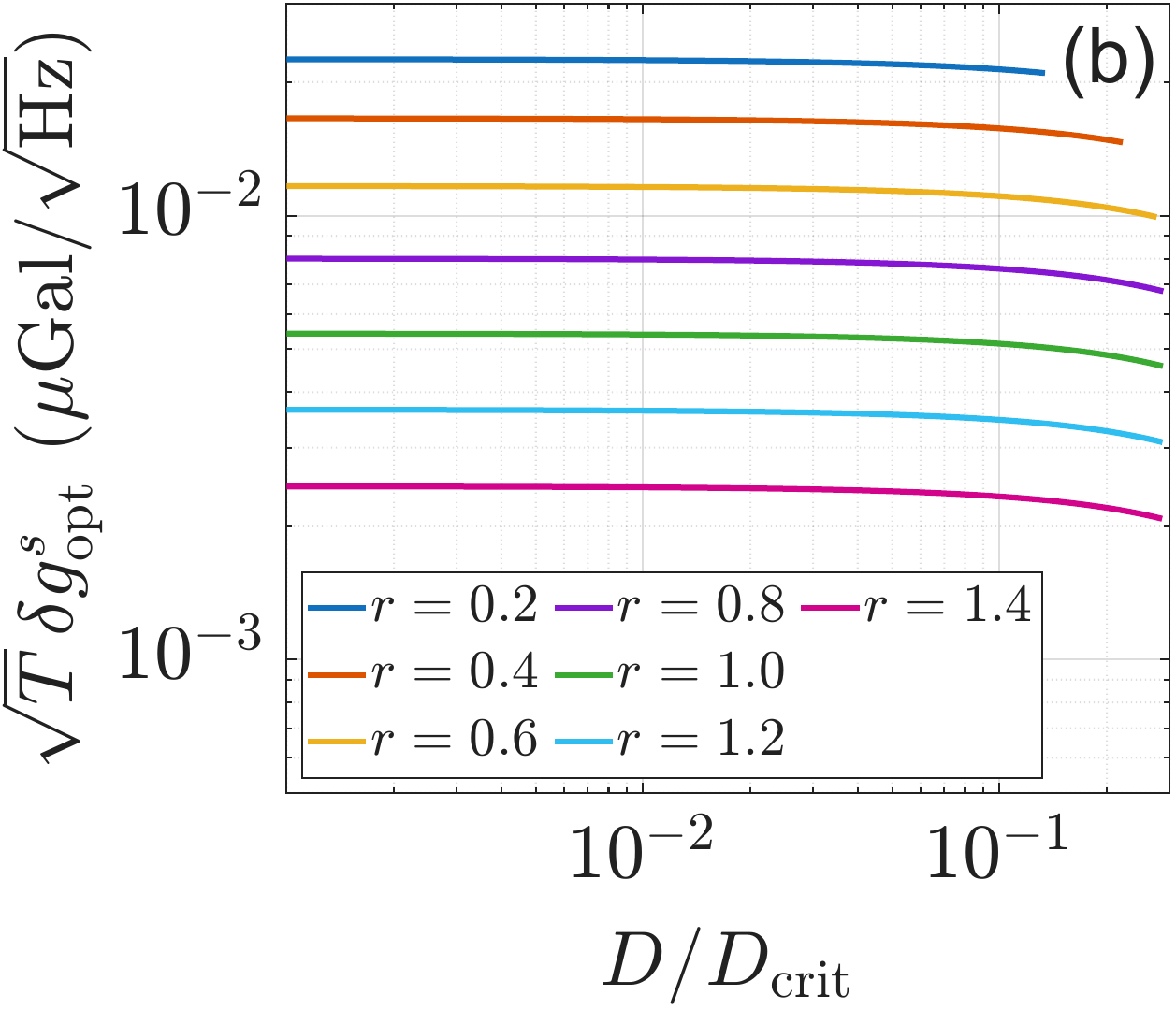}
    \includegraphics[width=0.49\linewidth]{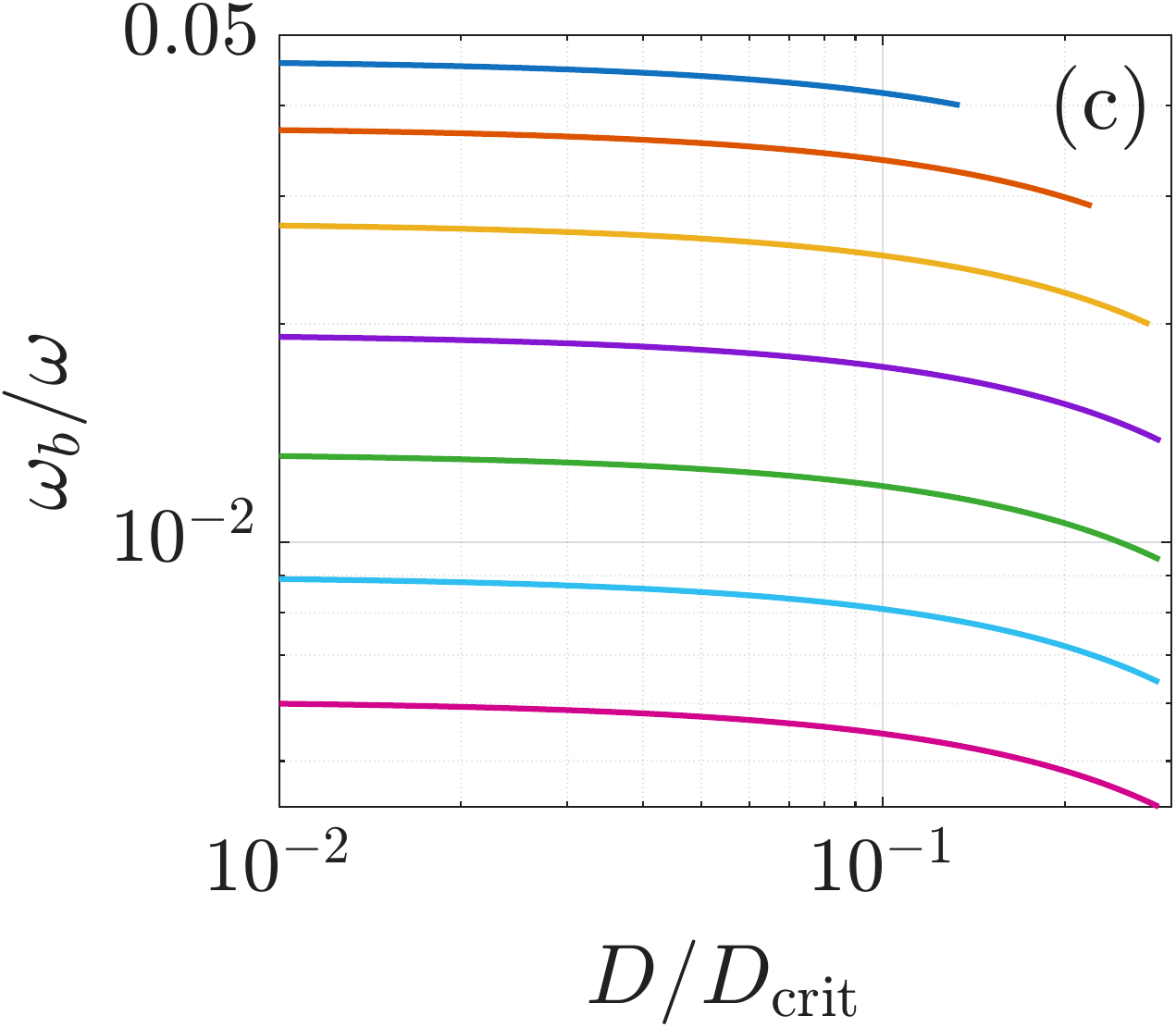}
    \includegraphics[width=0.49\linewidth]{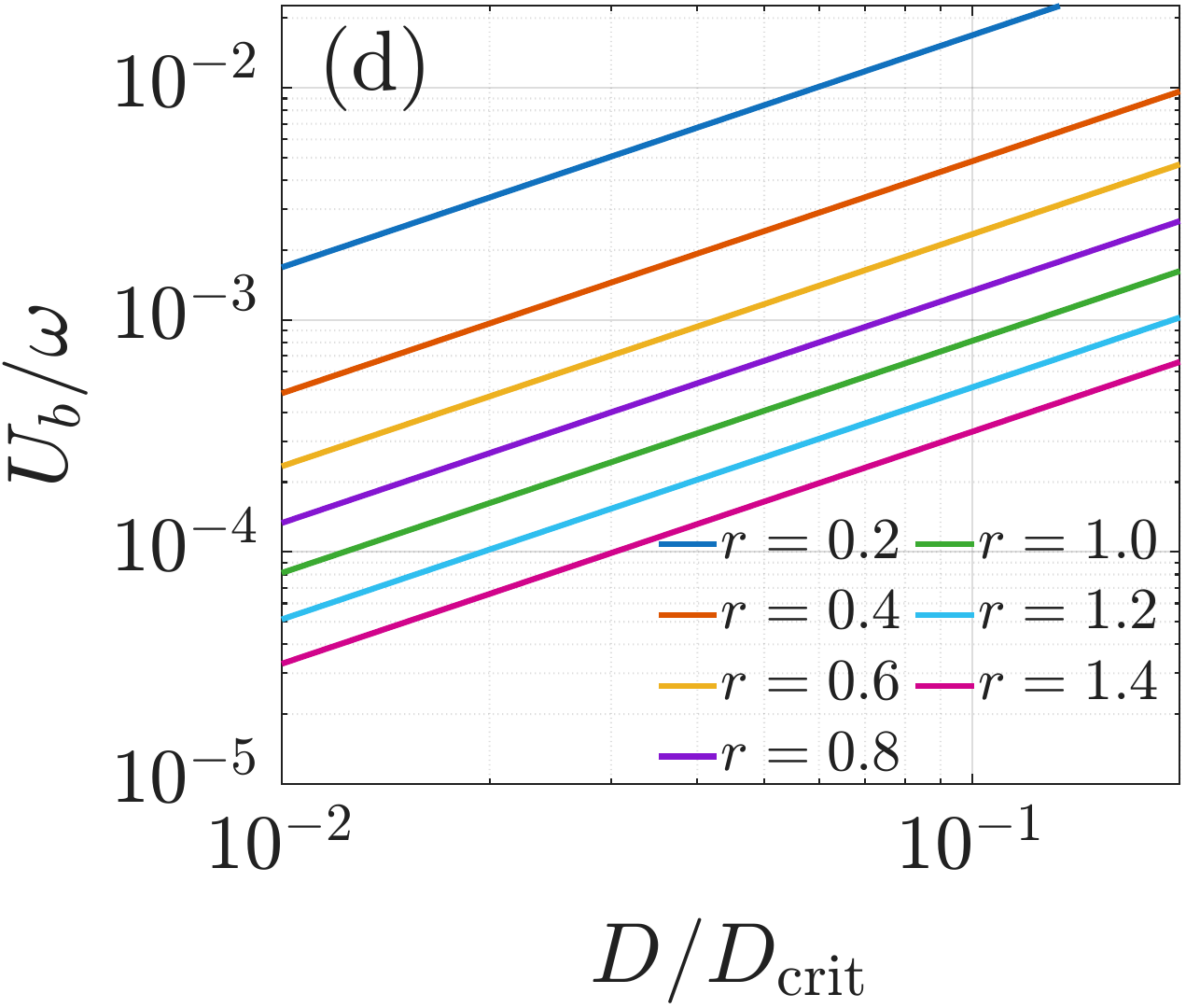}
    \caption{Coherent performance of the MSFQ gravimeter. 
(a)~Time-normalized sensitivity $\sqrt{T}\,\delta g^s_{\rm opt}$ 
as a function of $r$ at fixed $D/D_{\rm crit}=0.2$. 
The horizontal lines mark the MQ sensitivity 
$\sqrt{\pi\hbar\omega^2/8m}$ and the MCQ sensitivity 
$\sqrt{\pi\hbar\omega^2/16mN}$ for $N{=}10$ reported in Ref.~\cite{Huo2026QuantumGravimetryMechanicalQubits}. 
(b)~$\sqrt{T}\,\delta g^s_{\rm opt}$ as a function of 
$D/D_{\rm crit}$ for various $r$'s.
(c)~Qubit frequency $\omega_b/\omega$ as a function of
$D/D_\text{crit}$ for the same values of $r$ as in panel~(b).
(d)~Effective anharmonicity $U_b/\omega$ as a function of
$D/D_\text{crit}$ for various $r$'s.
In both panels, each curve is shown only within the RWA-valid region
(see Appendix~\ref{app:Effective Hamiltonian}).
The physical parameters are $m{=}10^{-9}$~kg and $\omega/2\pi{=}1$~kHz, 
with $\delta/\omega{=}0.05$ throughout.}
    \label{fig:1}
\end{figure}
Fig.~\ref{fig:1} summarizes the coherent operating principle of the MSFQ gravimeter within the valid regions. 
Fig.~\ref{fig:1}(a) shows the time-normalized sensitivity 
$\sqrt{T}\,\delta g^s_{\rm opt}$ as a function of the squeezing parameter 
$r$ at fixed $D/D_{\rm crit}=0.2$ as a representative operating point that remains in the RWA-valid region while giving an appreciable reduction of $\omega_b$.
Obviously $\sqrt{T}\,\delta g^s_{\rm opt}$ decreases monotonically with $r$, reflecting the 
exponential enhancement $e^{r}$ of the gravity-induced coupling 
at the Hamiltonian level. 
For sufficiently large $r$, MSFQ suppresses both MQ benchmark 
$\sqrt{\pi\hbar\omega^2/8m}$ 
and MCQ benchmark $\sqrt{\pi\hbar\omega^2/16mN}$ with $N=10$, reported in Ref.~\cite{Huo2026QuantumGravimetryMechanicalQubits}, demonstrating that squeezing alone is sufficient to achieve a competitive sensitivity.
Fig.~\ref{fig:1}(b) shows the sensitivity as a function 
of $D/D_{\rm crit}$ for fixed values of $r\in\{0.2,\ldots,1.4\}$.
Two effects are simultaneously visible.
First, for any fixed $r$, the sensitivity improves monotonically as 
$D/D_{\rm crit}$ increases, because $\omega_b$ decreases 
and the coherent bound tightens accordingly.
Second, at any fixed $D/D_{\rm crit}$, increasing $r$ shifts the entire 
curve downward, confirming that squeezing is the dominant resource.
In Fig.~\ref{fig:1}(b), the plotted
range is set by the RWA boundary discussed in Appendix~\ref{app:Effective Hamiltonian}, and the curves are truncated at that boundary.
Figs.~\ref{fig:1}(c) and~(d) show,
respectively, the squeezed qubit frequency $\omega_b/\omega$ and
the effective anharmonicity $U_b/\omega$ as functions of
$D/D_\text{crit}$ for various $r$'s.
For any fixed $r$, increasing $D/D_\text{crit}$ simultaneously
reduces $\omega_b/\omega$ and raises $U_b/\omega$, since both
quantities are controlled by duffing parameter.
Larger $r$ shifts the $\omega_b/\omega$ curves in panel~(c) downward
and the $U_b/\omega$ curves in panel~(d) upward, amplifying both
effects simultaneously; it also compresses the RWA-valid range to
smaller $D/D_\text{crit}$.
This joint behavior is doubly favorable: a smaller $\omega_b$
lengthens the optimal interrogation time $t_\text{opt}=\pi/\omega_b$
and tightens the coherent sensitivity bound, while a larger $U_b$
exponentially increases the anharmonic protection of the
squeezed-qubit subspace and suppresses leakage to higher
squeezed-Fock levels.

\section{Effect of decoherence on quantum gravimeter}
The ideal results above assume that the MSFQ evolves coherently. We now include the dominant mechanical damping channel and derive the corresponding gravimetric sensitivity. In the laboratory frame, zero-temperature mechanical damping is described by~\cite{GardinerZoller2004QuantumNoise,BreuerPetruccione2002OpenQuantumSystems,Aspelmeyer2014CavityOptomechanics,Clerk2010QuantumNoise,Fazio2025ManyBodyOpenQuantumSystems}
\begin{equation}
\dot\rho
=
-\frac{i}{\hbar}[\hat H,\rho]
+
\frac{\gamma_0}{2}\mathcal D_{\hat a,\hat a^\dagger}[\rho],
\qquad
\mathcal D_{\hat A,\hat B}[\rho]
=
2\hat A\rho\hat B-\hat B\hat A\rho-\rho\hat B\hat A,
\end{equation}
where $\gamma_0$ is the mechanical energy-relaxation rate. Using the Bogoliubov transformation maps and projecting onto the squeezed-qubit subspace $\{|0\rangle_{\rm s},|1\rangle_{\rm s}\}$, results in the effective jump operator as $\hat L_{\rm s} = \cosh r\,\hat\sigma_-+ \sinh r\,\hat\sigma_+$.
Then decoherent squeezed-qubit gravimeter is governed by
\begin{equation}
\dot\rho
=
-\frac{i}{2}
\left[
\hat H,\rho
\right]
+
\frac{\gamma_0}{2}\mathcal D_{\hat L_{\rm s},\hat L_{\rm s}^\dagger}[\rho].
\label{eq:decoherent_ME_main}
\end{equation}
The Bloch vector $\bm v = (x, y, z)^T$ coresponding to $\rho$ evolves as
\begin{equation}
\dot{\bm v} = \mathsf{A}(\Omega)\,\bm v + \bm b,
\end{equation}
with the drift matrix
\begin{equation}
\mathsf{A} = 
\begin{pmatrix} 
-\Gamma_x & \omega_b & 0 \\ 
-\omega_b & -\Gamma_y & \Omega^{\rm s}_{\rm g} \\ 
0 & -\Omega^{\rm s}_{\rm g} & -\Gamma_z 
\end{pmatrix}, 
\qquad 
\bm b = \begin{pmatrix}0\\0\\-\gamma_0\end{pmatrix},
\end{equation}
and the three anisotropic decoherence rates
\begin{equation}
\Gamma_x = \tfrac{\gamma_0}{2}e^{-2r}, 
\qquad 
\Gamma_y = \tfrac{\gamma_0}{2}e^{2r}, 
\qquad 
\Gamma_z = \gamma_0\cosh(2r).
\end{equation}
Clearly, the same squeezing that amplifies the gravitational signal also anisotropically amplifies the noise.
The QFI is then
\begin{equation}
\mathcal{F}_Q^{\rm dec}(g) 
= 
\kappa_g^2\left[|\bm u|^2 + \frac{(\bm v \cdot \bm u)^2}{1 - |\bm v|^2}\right], 
\qquad 
\kappa_g^2 = \frac{2me^{2r}}{\hbar\omega},
\end{equation}
where $\bm u {=} \partial_\Omega \bm v$ is the sensitivity vector satisfying 
the same drift matrix, sourced by $\partial_\Omega \mathsf{A}$ (see Appendix~\ref{app:decoherence_fisher}).
The anisotropic damping redistributes information about $g$ among all three Bloch-vector components. Therefore, the standard population measurement is no longer optimal in the presence of decoherence. The optimized readout, obtained by measuring along the symmetric-logarithmic-derivative direction $\bm n_{\rm opt}$, extracts the full information stored in the Bloch vector. As a result, the optimized CFI coincides with the QFI at the chosen operating point (see Appendix~\ref{app:decoherence_fisher} and Fig.~\ref{fig:nopt_angles_app}).
\begin{figure}[t!]
    \centering
    \includegraphics[width=0.49\linewidth]{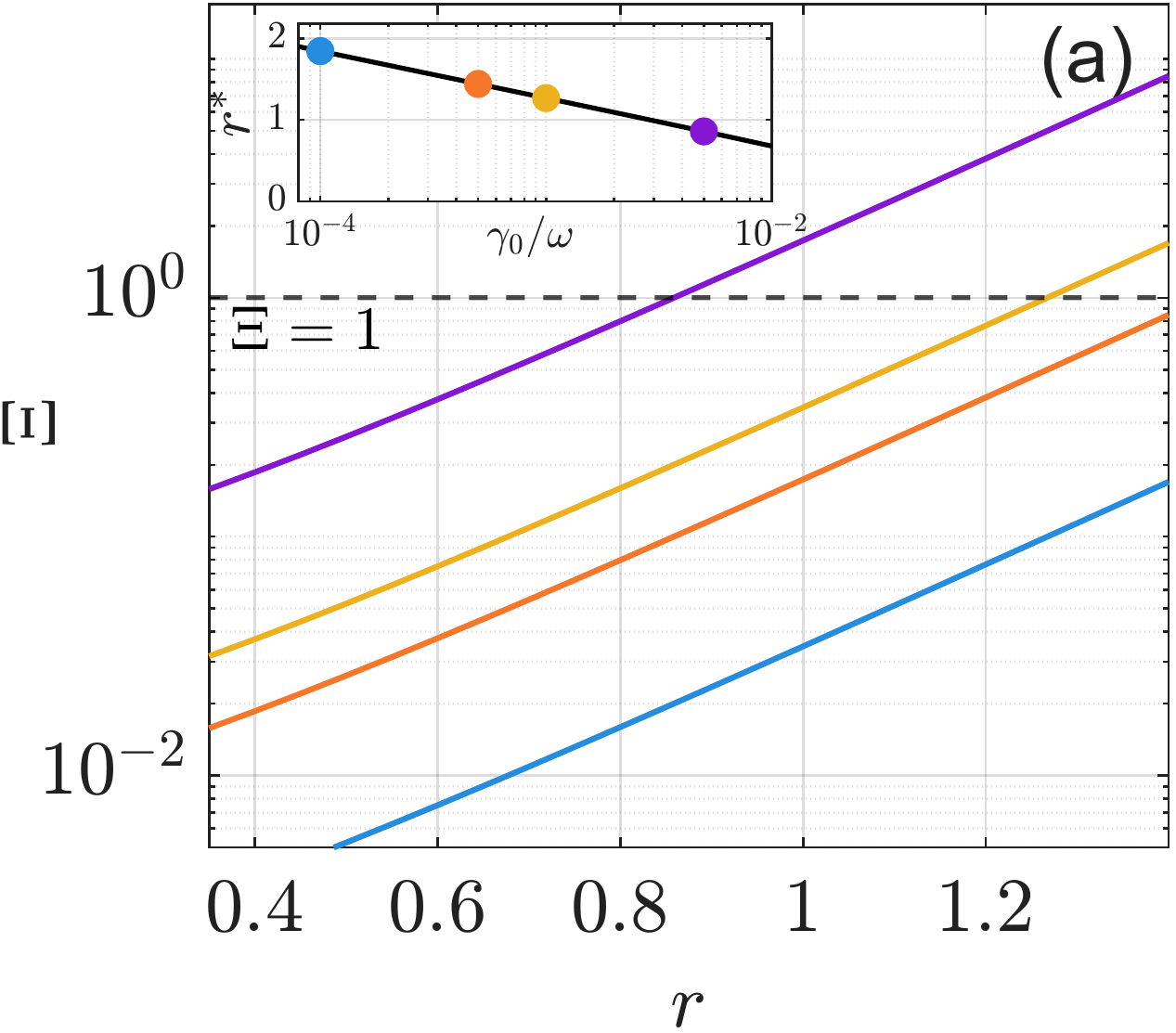}
    \includegraphics[width=0.49\linewidth]{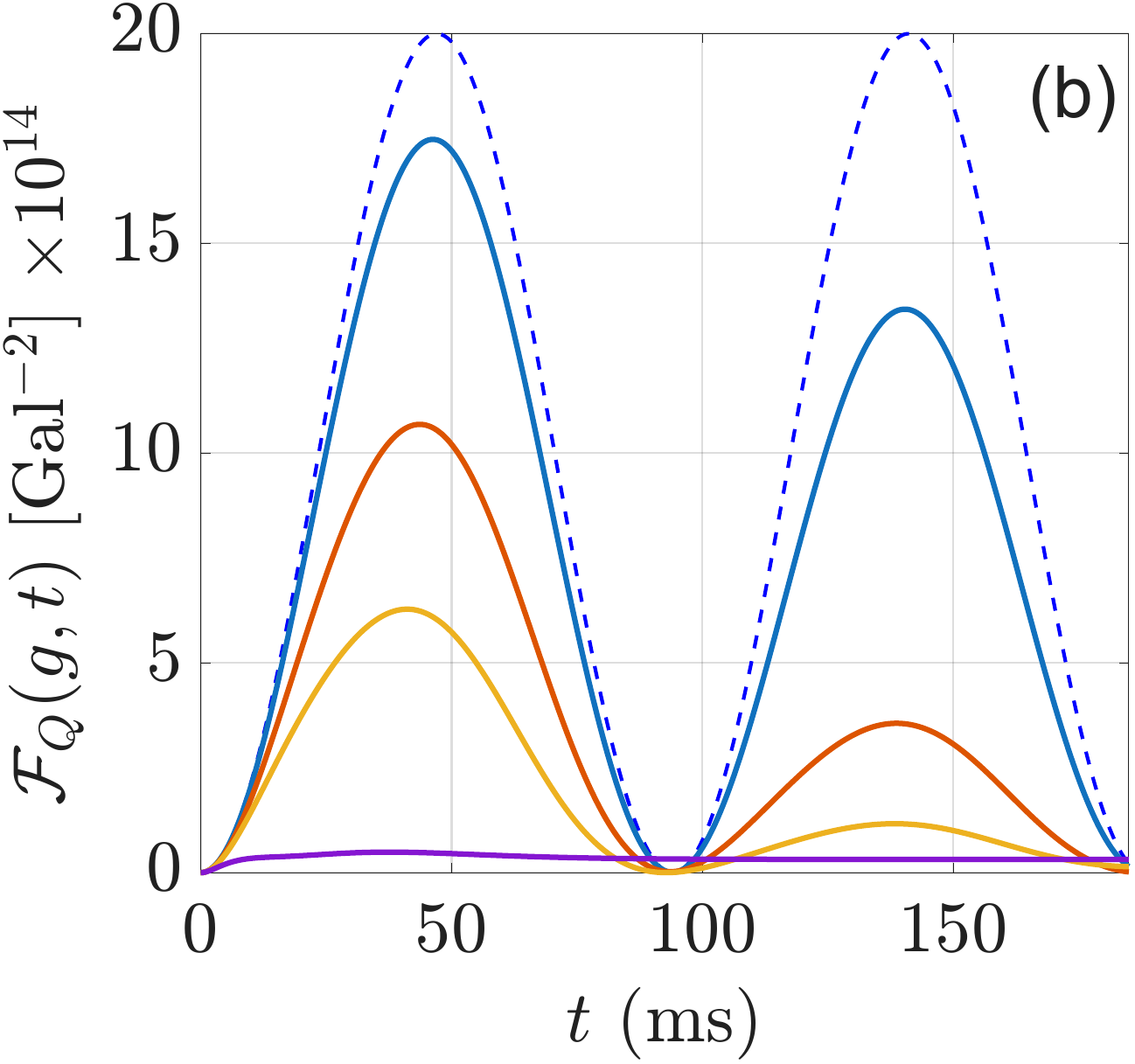}
    \includegraphics[width=0.49\linewidth]{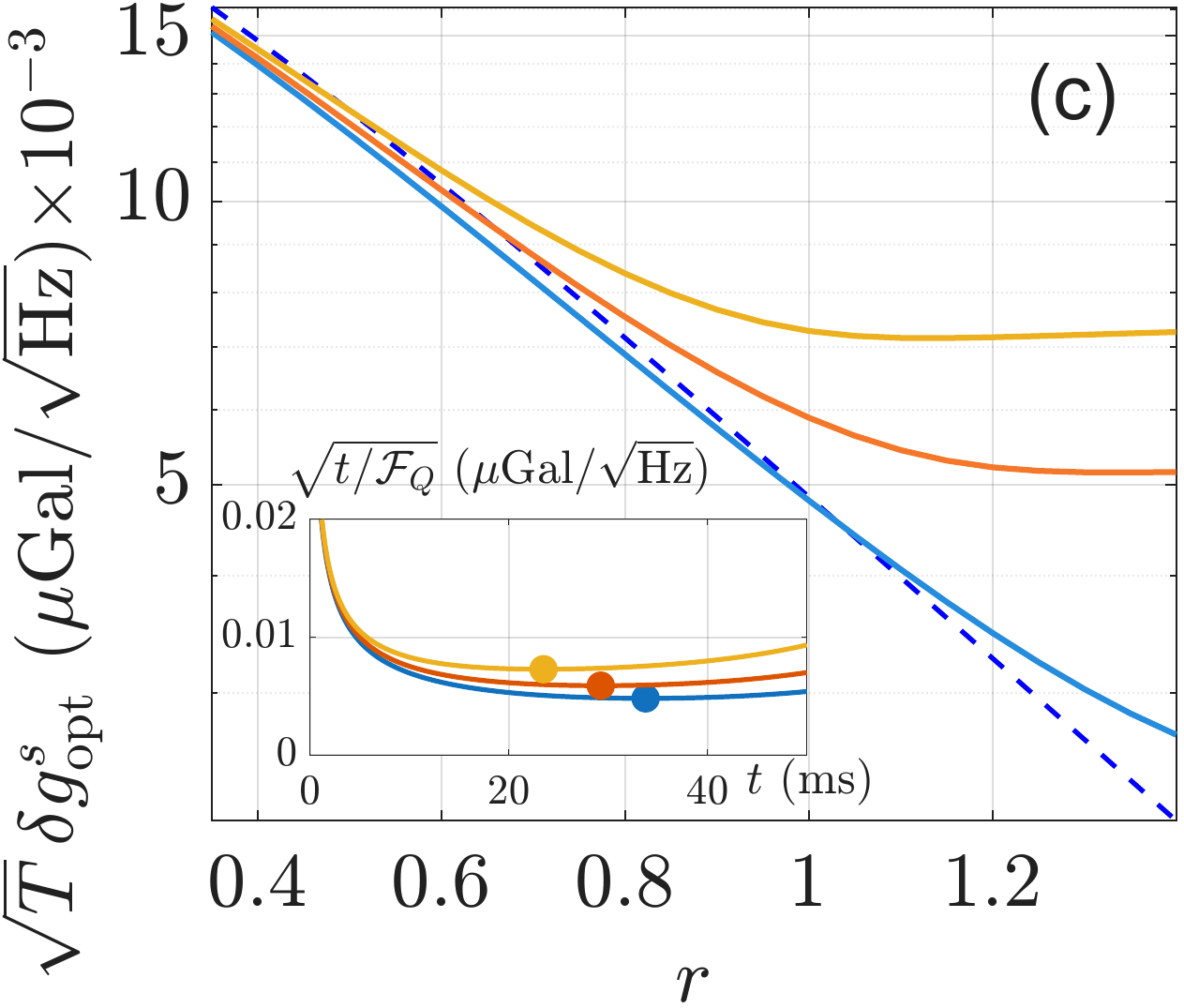}
    \includegraphics[width=0.49\linewidth]{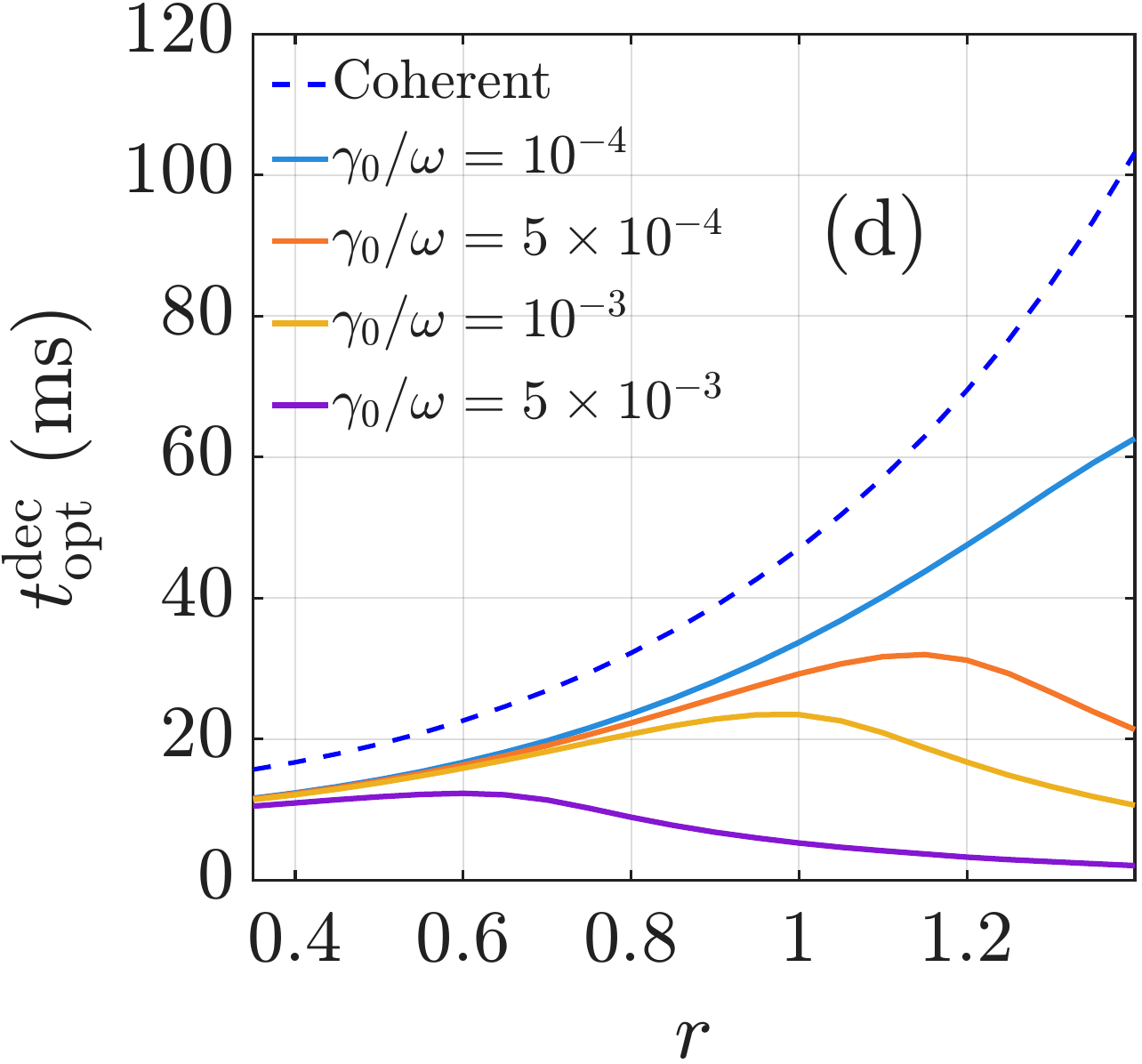}
    \caption{Decoherent performance of the MSFQ gravimeter, for $m{=}10^{-9}$~kg, $\omega/2\pi{=}1$~kHz,
    and $\delta/\omega{=}0.05$ throughout.
    (a)~Competition ratio $\Xi{\equiv}\Gamma_{\rm eff}/\omega_b$
    as a function of $r$ at fixed $D/D_{\rm crit}{=}0.2$,
    for various $\gamma_0$'s.
    The dashed line $\Xi{=}1$ marks the crossover between the
    coherent ($\Xi\ll 1$) and decoherent ($\Xi\gg 1$) regimes.
    (b)~QFI $\mathcal{F}_Q(g,t)$ as a function of time at fixed
    $D/D_{\rm crit}{=}0.2$ and $r{=}1$, for various $\gamma_0$'s
    and the coherent reference ($\gamma_0{=}0$,
    dashed).
    (c)~Optimized sensitivity $\sqrt{T}\,\delta g^s_{\rm opt}$
    as a function of $r$ at fixed $D/D_{\rm crit}{=}0.2$,
    evaluated at the decoherence-limited interrogation time
    $t_{\rm opt}^{\rm dec}{\equiv}\arg\min_t\sqrt{t/\mathcal{F}_Q(g,t)}$ shown as markers in the inset.
    (d)~Decoherence-limited optimal interrogation time $t_{\rm opt}^{\rm dec}$
    as a function of $r$ at fixed $D/D_{\rm crit}{=}0.2$, for various
    $\gamma_0$'s with the coherent reference
    $t_{\rm opt}{=}\pi/\omega_b$ (dashed). Note that the considered $D/D_{\rm crit}{=}0.2$ lies inside the safe RWA region for the plotted values of $r$.}
    \label{fig:Noisy}
\end{figure}
\\ \\
The decoherent QFI is governed by the prefactor $\kappa_g^2$ and by the sensitivity vector $\bm u(t) {=} \partial_\Omega\bm v(t)$, whose magnitude and orientation are set by the competition between coherent Rabi precession at frequency $\omega_b$ and the anisotropic decay channel. This competition is governed by 
\begin{equation}
\Xi \equiv \frac{\Gamma_{\rm eff}}{\omega_b},
\qquad
\Gamma_{\rm eff} \approx \frac{\gamma_0}{2}e^{2r},
\label{eq:xi}
\end{equation}
where $\Gamma_{\rm eff}$ is set by the dominant decoherence rate $\Gamma_y$.
Fig.~\ref{fig:Noisy}(a) shows $\Xi$ as a function of $r$ at fixed
$D/D_{\rm crit}{=}0.2$, for four values of $\gamma_0/\omega$.
For any fixed $\gamma_0$, $\Xi$ grows with $r$ exponentially.
Each curve crosses the threshold $\Xi{=}1$ at a squeezing value
$r^*(\gamma_0)$ that shifts to smaller $r$ as $\gamma_0$ increases, see the inset of Fig.~\ref{fig:Noisy}(a).
Below this threshold, the qubit completes a Rabi oscillation before
decoherence acts, and the coherent regime is recovered; above it, the
sensor relaxes before the gravity signal is encoded, as it can be seen from 
Fig.~\ref{fig:Noisy}(b). This panel reports the QFI dynamics at fixed $r{=}1$ and
$D/D_{\rm crit}{=}0.2$ for the same set of $\gamma_0/\omega$ values,
together with the coherent reference.
In the coherent limit ($\gamma_0{=}0$), the QFI oscillates
periodically with maxima at odd multiples of $t_{\rm opt}{=}\pi/\omega_b$,
as predicted by Eq.~\eqref{eq:FQ_g_weak_x0_main}.
For $\gamma_0/\omega{=}10^{-4}$, the system lies below $\Xi{=}1$ at
$r{=}1$ and the envelope of the oscillations is only weakly damped,
so the first peak closely approaches the coherent value.
As $\gamma_0$ increases and $\Xi$ exceeds unity, the QFI peak is
progressively suppressed and displaced to earlier times, narrowing
the window of useful sensing.
See the results for $\gamma_0/\omega{=}5\times10^{-3}$ in this panel. 
Fig.~\ref{fig:Noisy}(c) shows the optimized
sensitivity $\sqrt{T}\,\delta g^s_{\rm opt}$ as a function of $r$
at fixed $D/D_{\rm crit}{=}0.2$, evaluated at the decoherence-limited
optimal time $t_{\rm opt}^{\rm dec}{\equiv}\arg\min_t\sqrt{t/\mathcal{F}_Q(g,t)}$
(markers in inset).
For $\gamma_0/\omega{=}10^{-4}$, the decoherent curve follows the
coherent reference closely up to $r\approx r^*$, beyond which the
sensitivity saturates rather than improving further.
Larger $\gamma_0$ brings this saturation to smaller $r$, consistent with the crossover values identified
in panel~(a). Finally, Fig.~\ref{fig:Noisy}(d) shows extracted $t_{\rm opt}^{\rm dec}$ as a
function of $r$ for the same $\gamma_0$ values, with the coherent
reference $t_{\rm opt}{=}\pi/\omega_b$ (dashed).
The coherent time grows with $r$ since $\omega_b$ decreases with
squeezing.
In the presence of damping, $t_{\rm opt}^{\rm dec}$ is pulled below
this reference at all $r$, and the gap widens as either $r$ or
$\gamma_0$ increases.
Together, panels~(a)--(d) identify the practical operating window
for the MSFQ gravimeter: increasing $r$  raises $\Xi$, so the optimal squeezing
for a given $\gamma_0$ is set by the condition $\Xi\lesssim 1$,
i.e., $r\lesssim r^*(\gamma_0)$.
\\ \\ 
Having elucidated the influence of zero-temperature mechanical damping, we now consider the case of a finite-temperature environment. In this case, the mechanical mode undergoes both phonon emission and thermal excitation, resulting in the thermal occupation
\begin{equation}
n_{\rm th}=\frac{1}{e^{\hbar\omega/k_BT}-1},
\end{equation}
which renormalizes the anisotropic decoherence rates by the factor $2n_{\rm th}+1$, while leaving the sensing Hamiltonian and the gravity-induced coupling $\Omega_g^{\rm s}$ unchanged. Consequently, the competition between coherent signal accumulation and thermal decoherence is governed by
\begin{equation}
\Xi_T=\frac{\Gamma_{\rm eff}^{T}}{\omega_b},
\qquad
\Gamma_{\rm eff}^{T}\simeq
\frac{\gamma_0}{2}(2n_{\rm th}+1)e^{2r},
\end{equation}
and the coherent sensing regime is maintained whenever $\Xi_T{\lesssim}1$. A detailed numerical analysis at $T{=}10~{\rm mK}$ is presented in Appendix~C. We show that thermal noise reduces the maximum achievable QFI and shifts the optimal interrogation time to shorter values as either the squeezing parameter or the  damping increases. Nevertheless, for experimentally achievable ultrahigh-quality mechanical resonators~\cite{Gieseler2013NatPhys,Delic2020,Ranjit2016}, the degradation remains modest over a broad squeezing range, allowing the MSFQ gravimeter to retain most of its coherent sensitivity enhancement. These results demonstrate that the proposed sensing protocol remains robust against realistic finite-temperature decoherence and is compatible with the operating conditions of current levitated optomechanical platforms.

\section{Conclusion}
\label{sec:conclusion}

In this paper, we propose a mechanical squeezed-Fock gravimeter in which gravity is sensed directly through the CM motion of a levitated mesoscopic particle. The protocol combines the favorable mass scaling of quantized mechanical motion with the quadrature amplification generated by a detuned two-phonon pump. In the squeezed-Fock basis, the gravitational force couples to the anti-squeezed quadrature, enhancing the effective gravity-induced transition rate by a factor $e^r$ while preserving the direct mechanical coupling to the particle mass.
In the coherent regime and in the weak-force limit, the exact QFI results in $\sqrt{T}\delta g_{\rm opt}^{\rm s}
\simeq e^{-r}\sqrt{\pi\hbar\omega\omega_b/8m}.$
This expression identifies the key resources of the sensor: squeezing-enhanced force coupling, large mechanical mass, and a tunably small qubit splitting $\omega_b$. Increasing Duffing nonlinearity towards its allowed maximum improves sensitivity by reducing $\omega_b$ and simultaneously strengthens the anharmonic protection of the squeezed-qubit subspace.
Although we use a Duffing nonlinearity as the microscopic model, the gravimetric mechanism does not 
rely on the microscopic origin of the anharmonicity. What is required is an effective squeezed-mode 
Hamiltonian with a resolvable two-level subspace and a linear displacement coupling to the 
gravitational force. Therefore, any mechanism that produces a stable Kerr-type squeezed-mode spectrum 
with leakage rates low compared with the effective anharmonicity can realize the same sensing 
principle. In such a case, the signal enhancement $\Omega_g^s\propto e^r$ follows from the quadrature 
transformation and is unchanged, while the quantitative expressions for $\omega_b$, $U_b$, $D_{\rm 
crit}$, and the RWA boundary must be recalculated for the specific anharmonic mechanism.
We also showed that population measurement in the squeezed-Fock basis saturates the QFI.
We further analyzed the dissipative dynamics, for both zero- and finite-temperature baths, and showed that squeezing converts ordinary mechanical damping into anisotropic qubit noise. As a result, the optimal sensitivity is governed by a trade-off between signal amplification and decoherence. When the effective decoherence rate remains comparable with $\omega_b$, the coherent advantage is largely retained; otherwise, the QFI is suppressed and the optimal interrogation time shifts to shorter values. In this noisy regime, the optimal measurement is determined by the SLD direction.
Overall, our results establish the MSFQ as a promising platform for quantum gravimetry. The sensor preserves the direct CM coupling to gravity, enhances the signal through squeezing, and provides a clear operating window set by the balance between squeezing, anharmonic protection, interrogation time, and decoherence.

\appendix
\makeatletter
\@addtoreset{equation}{section}
\@addtoreset{figure}{section}
\@addtoreset{table}{section}
\makeatother
\renewcommand{\theequation}{\Alph{section}\arabic{equation}}
\renewcommand{\thefigure}{\Alph{section}\arabic{figure}}
\renewcommand{\thetable}{\Alph{section}\arabic{table}}
\renewcommand{\theHequation}{\Alph{section}.\arabic{equation}}
\renewcommand{\theHfigure}{\Alph{section}.\arabic{figure}}
\renewcommand{\theHtable}{\Alph{section}.\arabic{table}}
\begin{widetext}
\section{Effective Hamiltonian}
\label{app:Effective Hamiltonian}
This appendix derives the effective Hamiltonian used in the main text. Starting from 
\begin{align}
\hat H_{\rm rot}
 &{=}
\hbar \delta \, \hat a^\dagger \hat a
-
\hbar D \hat a^{\dagger}\hat a^{\dagger}\hat a \hat a
+
\frac{\hbar A_p}{2}
\left(
e^{-i\theta}\hat a^2
+
e^{i\theta}\hat a^{\dagger 2}
\right)
\end{align}
we substitute the Bogoliubov transformation $\hat a {=} c\,\hat b - e^{i\theta}s\,\hat b^\dagger, $ and $\hat a^\dagger {=} c\,\hat b^\dagger - e^{-i\theta}s\,\hat b$ in which $c \equiv \cosh r,$ and $s \equiv \sinh r,$ 
to simplify the Hamiltonian.
First,
\begin{align}
\hat a^\dagger \hat a
{=}
\left(c\hat b^\dagger - e^{-i\theta}s\hat b\right)
\left(c\hat b - e^{i\theta}s\hat b^\dagger\right)
{=}
c^2 \hat b^\dagger \hat b
- cs\,e^{i\theta}\hat b^\dagger \hat b^\dagger
- cs\,e^{-i\theta}\hat b \hat b
+ s^2 \hat b \hat b^\dagger
{=}
(c^2+s^2)\hat b^\dagger \hat b
+s^2
-cs\left(e^{i\theta}\hat b^{\dagger 2}+e^{-i\theta}\hat b^2\right),
\label{eq:adag_a}
\end{align}
where we used $\hat b\hat b^\dagger{=}\hat b^\dagger \hat b+1$.
Next,
\begin{align}
\hat a^2
{=}
\left(c\hat b - e^{i\theta}s\hat b^\dagger\right)^2
{=}
c^2 \hat b^2
-cs\,e^{i\theta}\left(\hat b\hat b^\dagger+\hat b^\dagger \hat b\right)
+s^2 e^{i2\theta}\hat b^{\dagger 2},
\end{align}
and similarly
\begin{align}
\hat a^{\dagger 2}
&{=}
c^2 \hat b^{\dagger 2}
-cs\,e^{-i\theta}\left(\hat b^\dagger \hat b+\hat b\hat b^\dagger\right)
+s^2 e^{-i2\theta}\hat b^2.
\end{align}
Therefore,
\begin{align}
e^{-i\theta}\hat a^2 + e^{i\theta}\hat a^{\dagger 2}
&{=}
(c^2+s^2)\left(e^{-i\theta}\hat b^2+e^{i\theta}\hat b^{\dagger 2}\right)
-2cs\left(2\hat b^\dagger \hat b+1\right).
\label{eq:parametric_part}
\end{align}
Substituting Eqs.~\eqref{eq:adag_a} and \eqref{eq:parametric_part} into the quadratic part of the Hamiltonian $\hat H_{\rm quad}{=}
\hbar \delta \hat a^\dagger \hat a +
\frac{\hbar A_p}{2}
\left( e^{-i\theta}\hat a^2+e^{i\theta}\hat a^{\dagger 2} \right) $, we obtain
\begin{align}
\hat H_{\rm quad} {=} 
\hbar\Big[\delta(c^2+s^2)-2A_p cs\Big]\hat b^\dagger \hat b
\quad
+\hbar\Big[\frac{A_p}{2}(c^2+s^2)-\delta cs\Big]
\left(e^{-i\theta}\hat b^2+e^{i\theta}\hat b^{\dagger 2}\right)
\quad
+\hbar\Big(\delta s^2-A_p cs\Big).
\label{eq:Hquad_before_choice}
\end{align}
Using $c^2+s^2{=}\cosh 2r$ and $2cs{=}\sinh 2r$, 
Eq.~\eqref{eq:Hquad_before_choice} becomes
\begin{align}
\hat H_{\rm quad}
{=}
\hbar\Big(\delta\cosh 2r-A_p \sinh 2r\Big)\hat b^\dagger \hat b
\quad
+\frac{\hbar}{2}\Big(A_p \cosh 2r-\delta \sinh 2r\Big)
\left(e^{-i\theta}\hat b^2+e^{i\theta}\hat b^{\dagger 2}\right)
+\text{const.}
\label{eq:Hquad_compact}
\end{align}
We choose the squeezing parameter \(r\) such that the anomalous terms vanish:
\begin{equation}
A_p \cosh 2r-\delta \sinh 2r {=} 0
\qquad \Longrightarrow \qquad
\tanh 2r {=} \frac{A_p}{\delta}.
\label{eq:tanh_condition}
\end{equation}
With this choice, $\delta_b \equiv \delta\cosh 2r-A_p \sinh 2r
{=}\sqrt{\delta^2-A_p^2},$
and thus $\hat H_{\rm quad} {=} \hbar \delta_b \hat b^\dagger \hat b +\text{const.}$
\\
For the Duffing term
\begin{equation}
\hat H_D
{=}
-\hbar D\,\hat a^{\dagger 2}\hat a^2,
\end{equation}
one has $\hat a^{\dagger 2}\hat a^2
{=}
\left(c\hat b^\dagger-e^{-i\theta}s\hat b\right)^2
\left(c\hat b-e^{i\theta}s\hat b^\dagger\right)^2.$
After expanding this expression and repeatedly using the bosonic commutation relation
\(
[\hat b,\hat b^\dagger]{=}1
\)
to normal-order the operators, one finds
\begin{align}
\hat a^{\dagger 2}\hat a^2
&{=}
\left(8c^2s^2+4s^4\right)\hat b^\dagger \hat b
+
\frac{3\cosh 4r+1}{4}\hat b^{\dagger 2}\hat b^2
\nonumber\\
&\quad
+\frac{\sinh^2 2r}{4}
\left(
e^{i2\theta}\hat b^{\dagger 4}
+
e^{-i2\theta}\hat b^4
\right)
-\frac{\sinh 4r}{2}
\left(
e^{i\theta}\hat b^{\dagger 3}\hat b
+
e^{-i\theta}\hat b^\dagger \hat b^3
\right)
+\frac{\sinh 2r\left(3\cosh 2r-2\right)}{2}
\left(
e^{i\theta}\hat b^{\dagger 2}
+
e^{-i\theta}\hat b^2
\right)
+\text{const.}
\label{eq:quartic_normalordered}
\end{align}
Therefore,
\begin{align}
\hat H_D
&{=}
-\hbar D\left(8c^2s^2+4s^4\right)\hat b^\dagger \hat b
-\hbar D\,\frac{3\cosh 4r+1}{4}\hat b^{\dagger 2}\hat b^2
\nonumber\\
&\quad
-\hbar D\,\frac{\sinh^2 2r}{4}
\left(
e^{i2\theta}\hat b^{\dagger 4}
+
e^{-i2\theta}\hat b^4
\right)
+\hbar D\,\frac{\sinh 4r}{2}
\left(
e^{i\theta}\hat b^{\dagger 3}\hat b
+
e^{-i\theta}\hat b^\dagger \hat b^3
\right)
-\hbar D\,\frac{\sinh 2r\left(3\cosh 2r-2\right)}{2}
\left(
e^{i\theta}\hat b^{\dagger 2}
+
e^{-i\theta}\hat b^2
\right)
+\text{const.}
\label{eq:HD_full}
\end{align}
Collecting all the above equations and dropping constant energy shifts, we get
\begin{align}
\hat H_{\rm rot}
&{=}
\hbar \omega_b \hat b^\dagger \hat b
-\hbar U_b \hat b^{\dagger 2}\hat b^2
\nonumber\\
&\quad
-\hbar D\,\frac{\sinh^2 2r}{4}
\left(
e^{i2\theta}\hat b^{\dagger 4}
+
e^{-i2\theta}\hat b^4
\right)
+\hbar D\,\frac{\sinh 4r}{2}
\left(
e^{i\theta}\hat b^{\dagger 3}\hat b
+
e^{-i\theta}\hat b^\dagger \hat b^3
\right)
-\hbar D\,\frac{\sinh 2r\left(3\cosh 2r-2\right)}{2}
\left(
e^{i\theta}\hat b^{\dagger 2}
+
e^{-i\theta}\hat b^2
\right)
\label{eq:H_before_RWA}
\end{align}
where
\begin{equation}
\omega_b
{=}
\delta_b-D\left(8c^2s^2+4s^4\right)
{=}
\sqrt{\delta^2-A_p^2}
-D\left(8\cosh^2 r\,\sinh^2 r+4\sinh^4 r\right),
\label{eq:omega_b}
\end{equation}
and
\begin{equation}
U_b
{=}
\frac{D}{4}\left(3\cosh 4r+1\right).
\label{eq:Ub}
\end{equation}
To apply the rotating-wave approximation, we move to the interaction picture with respect to
\begin{equation}
\hat H_0{=}\hbar\omega_b \hat b^\dagger \hat b.
\end{equation}
Therefore, $\hat b(t){=}\hat b\,e^{-i\omega_b t},$ and $\hat b^\dagger(t){=}\hat b^\dagger e^{i\omega_b t}$.
This results in acquiring the rapid phases by the non-number-conserving quartic terms as
\begin{align}
\hat b^{\dagger 4} &\rightarrow \hat b^{\dagger 4}e^{i4\omega_b t},
&
\hat b^4 &\rightarrow \hat b^4 e^{-i4\omega_b t},
\nonumber \\ 
\hat b^{\dagger 3}\hat b &\rightarrow \hat b^{\dagger 3}\hat b\,e^{i2\omega_b t},
&
\hat b^\dagger \hat b^3 &\rightarrow \hat b^\dagger \hat b^3\,e^{-i2\omega_b t},
\nonumber \\
\hat b^{\dagger 2} &\rightarrow \hat b^{\dagger 2}e^{i2\omega_b t},
&
\hat b^2 &\rightarrow \hat b^2 e^{-i2\omega_b t}. \nonumber
\end{align}
When $D\,\frac{\sinh^2 2r}{4} \ll 4\omega_b,$ $D\,\frac{\sinh 4r}{2} \ll 2\omega_b,$ and $D\,\frac{\sinh 2r(3\cosh 2r-2)}{2} \ll 2\omega_b,$ the above terms oscillate rapidly and can be neglected. Under this rotating-wave approximation (RWA), the effective Hamiltonian in the Schr{\"o}dinger picture becomes
\begin{align}
\hat H_{\rm eff}
&\simeq
\hbar \omega_b \hat b^\dagger \hat b
-\hbar U_b \hat b^{\dagger }\hat b^{\dagger }\hat b\hat b
\label{eq:H_eff_after_RWA}
\end{align}
For each choice of the detuning \(\delta\) and pump amplitude \(A_p\), the squeezing parameter \(r\) is fixed by the ratio \(A_p/\delta\). Therefore, increasing \(A_p/\delta\) increases \(r\), which enhances the gravity-induced coupling. However, the same squeezing also reduces the effective qubit frequency \(\omega_b\). The shape of $\omega_b$ in Eq.~(\ref{eq:omega_b}) obligates us to determine the boundary at which the qubit gap closes, namely $\omega_b{=}0$. As such we introduce $D_{\rm crit}{=}\sqrt{\delta^2-A_p^2} \Big/ \left( 8\cosh^2 r \sinh^2 r + 4\sinh^4 r \right)$.
Exactly at \(D=D_{\rm crit}\), the optimal interrogation time becomes infinitely long.
To prevent this and the violation of the RWA, one should choose \(D\) safely below \(D_{\rm crit}\), where the squeezed-qubit picture remains valid and the sensing time is still finite.
Fig.~\ref{fig:FigA1} maps the RWA validity conditions
quantitatively. Fig.~\ref{fig:FigA1}(a) displays the largest of the three condition ratios in the
$(r,\,D/D_{\rm crit})$ plane. 
The RWA is valid in the green region and violated in the red region. As $r$ increases, the valid region
shrinks rapidly, because the dropped terms grow as $\sinh(2r)\sim e^{2r}$ while
$\omega_b$ simultaneously decreases.
The black line defines $D_{\rm RWA}(r,\varepsilon,A_p)<D_{\rm crit}$ for $\varepsilon=0.1$, when one enforce 
$D\frac{\sinh^2(2r)}{4} < \varepsilon\cdot 4\omega_b$, $\ D\frac{\sinh(4r)}{2} < \varepsilon\cdot 2\omega_b,$ and $\frac{D\sinh(2r)(3\cosh 2r-2)}{2}< \varepsilon\cdot 2\omega_b.$ 
\begin{figure}
    \centering
    \includegraphics[width=0.35\linewidth]{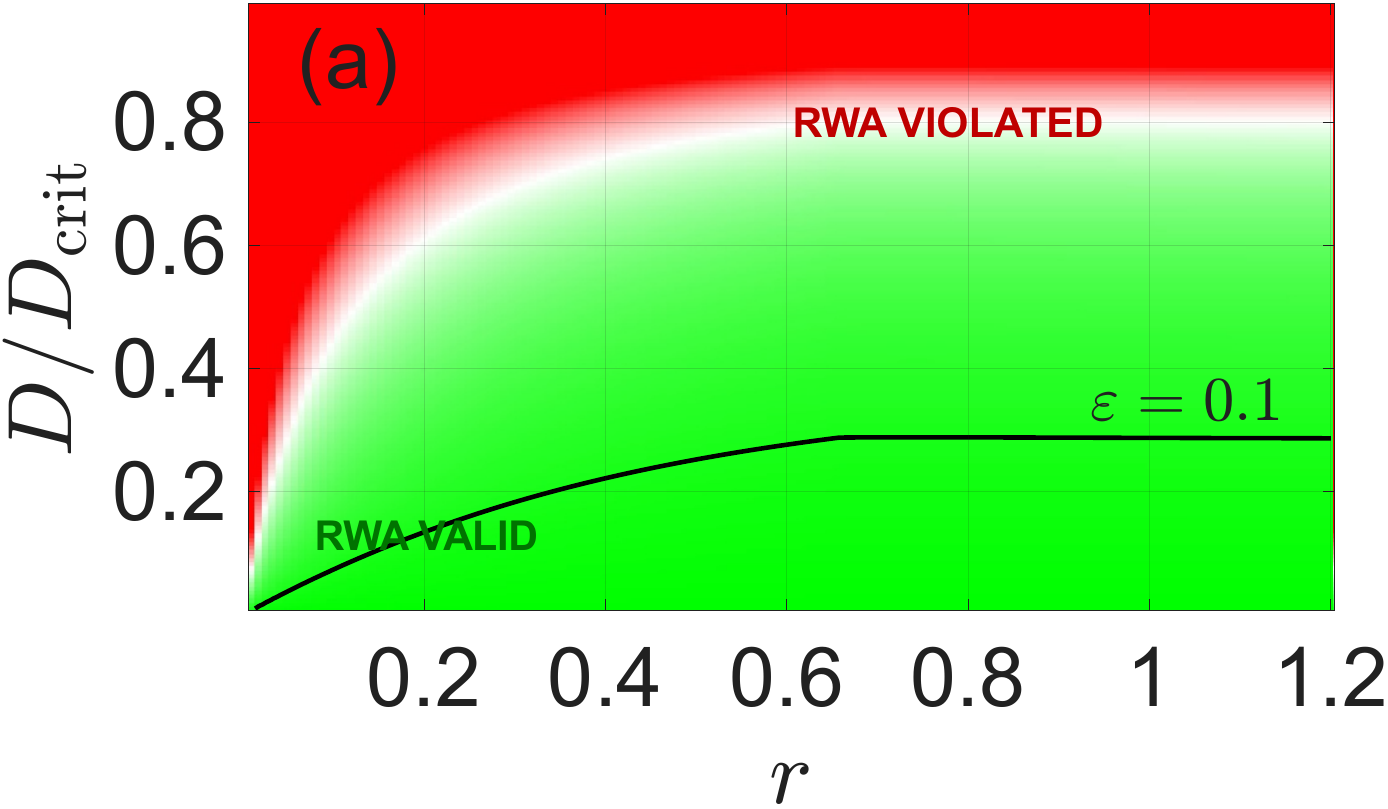}
    \includegraphics[width=0.35\linewidth]{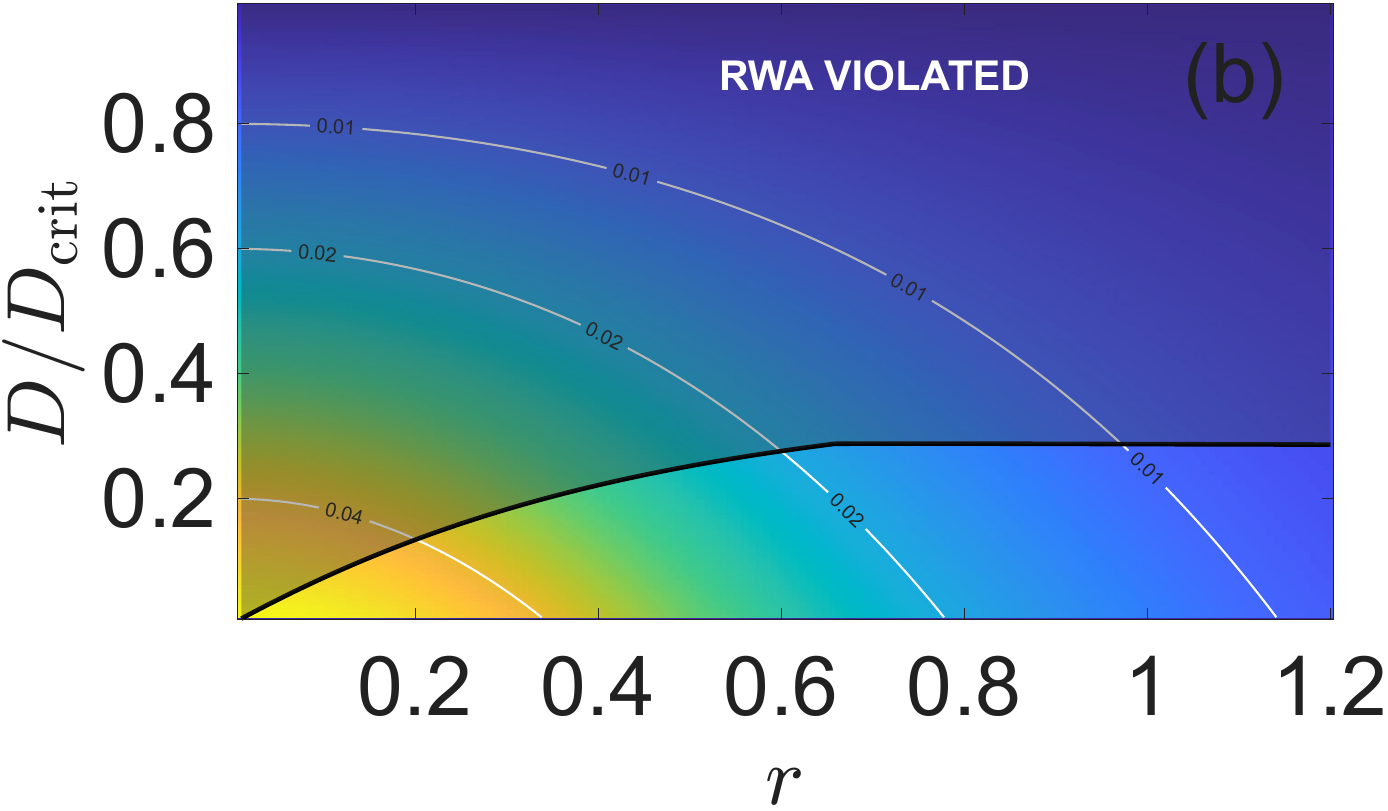}
    \caption{(a) RWA validity; Largest of the three RWA condition ratios,
$\max\!\left\{
\frac{D\sinh^{2}(2r)}{16\,\omega_{b}},\;
\frac{D\sinh(4r)}{4\,\omega_{b}},\;
\frac{D\sinh(2r)(3\cosh 2r-2)}{4\,\omega_{b}}
\right\}$,
in $(r,\,D/D_{\mathrm{crit}})$ plane.
The green (red) region marks where the RWA is valid (violated).
(b)~Rabi frequency $\omega_{b}/\omega$ in
$(r,\,D/D_{\mathrm{crit}})$ plane.
In both panels, the black contour defines the safe-operation boundary
$D_{\mathrm{RWA}}(r,\varepsilon,A_{p})$ for $\varepsilon=0.1$,
obtained by simultaneously enforcing
$D\sinh^{2}(2r)/4<\varepsilon\cdot 4\omega_{b}$,
$D\sinh(4r)/2<\varepsilon\cdot 2\omega_{b}$, and
$D\sinh(2r)(3\cosh 2r-2)/2<\varepsilon\cdot 2\omega_{b}$. Both panels use $\delta/\omega=0.05$.
}
    \label{fig:FigA1}
\end{figure}
Fig.~\ref{fig:FigA1}(b) maps the renormalized Rabi frequency
$\omega_{b}/\omega$ over the entire $(r,\,D/D_{\mathrm{crit}})$
plane. At fixed $r$, $\omega_{b}$ decreases monotonically with
$D/D_{\mathrm{crit}}$ and vanishes at the critical boundary.
At fixed $D/D_{\mathrm{crit}}$, increasing $r$ further suppresses
$\omega_{b}$ through the squeezing-induced renormalization. Therefore, one must choose $r$ and
$D$ small enough to keep the RWA valid and the
qubit gap open, yet large enough to benefit from squeezing-enhanced
signal amplification and reduced $\omega_{b}$.

\section{Coherent QFI and CFI}
\label{app:coherent_fisher}

In this section we derive the coherent Fisher information used in the main text. The effective squeezed-qubit Hamiltonian is
\begin{equation}
\hat H_{\rm eff}^{\rm s}
=
\frac{\hbar\omega_b}{2}\hat\sigma_z
+
\frac{\hbar\Omega_g^{\rm s}}{2}\hat\sigma_x .
\end{equation}
For compactness, we write
\begin{equation}
\Omega\equiv\Omega_g^{\rm s},
\qquad
\omega_{\rm R}=\sqrt{\omega_b^2+\Omega^2},
\qquad
\kappa_g\equiv\partial_g\Omega_g^{\rm s}
=
\frac{2mx_0e^r}{\hbar}.
\end{equation}
Starting from $|0\rangle_{\rm s}$, the evolved state is
\begin{equation}
|\psi(t)\rangle
=
\alpha(t)|0\rangle_{\rm s}+\beta(t)|1\rangle_{\rm s},
\end{equation}
with
\begin{align}
\alpha(t)
&=
\cos\left(\frac{\omega_{\rm R}t}{2}\right)
+i\frac{\omega_b}{\omega_{\rm R}}
\sin\left(\frac{\omega_{\rm R}t}{2}\right),\\
\beta(t)
&=
-i\frac{\Omega}{\omega_{\rm R}}
\sin\left(\frac{\omega_{\rm R}t}{2}\right).
\end{align}
The pure-state QFI is
\begin{equation}
\mathcal F_Q(g)
=
4\left[
\langle\partial_g\psi|\partial_g\psi\rangle
-
|\langle\psi|\partial_g\psi\rangle|^2
\right].
\end{equation}
Since $g$ enters only through $\Omega$, the chain rule gives
\begin{equation}
\mathcal F_Q(g)=\kappa_g^2\mathcal F_Q(\Omega).
\end{equation}
A direct calculation gives
\begin{align}
\mathcal F_Q(g)
&=
\kappa_g^2
\frac{
\Omega^4\omega_{\rm R}^2t^2
+2\Omega^2\omega_b^2\omega_{\rm R}t\sin(\omega_{\rm R}t)
+4\omega_b^2\omega_{\rm R}^2
\sin^2\left(\frac{\omega_{\rm R}t}{2}\right)
-\Omega^2\omega_b^2\sin^2(\omega_{\rm R}t)
}
{\omega_{\rm R}^6}.
\label{eq:FQ_g_exact_appendix}
\end{align}

The population of the excited squeezed-Fock state is
\begin{equation}
P_1(t)
=
\frac{\Omega^2}{\omega_{\rm R}^2}
\sin^2\left(\frac{\omega_{\rm R}t}{2}\right).
\end{equation}
The CFI associated with the binary outcomes $P_1$ and $P_0=1-P_1$ is
\begin{equation}
\mathcal F_C(g)
=
\frac{[\partial_gP_1(t)]^2}{P_1(t)[1-P_1(t)]}.
\end{equation}
Using
\begin{equation}
\partial_\Omega P_1
=
\frac{2\Omega\omega_b^2}{\omega_{\rm R}^4}
\sin^2\left(\frac{\omega_{\rm R}t}{2}\right)
+
\frac{\Omega^3t}{2\omega_{\rm R}^3}
\sin(\omega_{\rm R}t),
\end{equation}
one obtains
\begin{align}
\mathcal F_C(g)
&=
\kappa_g^2
\frac{
\left[
\dfrac{2\Omega\omega_b^2}{\omega_{\rm R}^4}
\sin^2\left(\dfrac{\omega_{\rm R}t}{2}\right)
+
\dfrac{\Omega^3t}{2\omega_{\rm R}^3}
\sin(\omega_{\rm R}t)
\right]^2
}
{
\dfrac{\Omega^2}{\omega_{\rm R}^2}
\sin^2\left(\dfrac{\omega_{\rm R}t}{2}\right)
\left[
1-
\dfrac{\Omega^2}{\omega_{\rm R}^2}
\sin^2\left(\dfrac{\omega_{\rm R}t}{2}\right)
\right]
}.
\label{eq:CFI_g_exact_appendix}
\end{align}
In the weak-force regime $\Omega\ll\omega_b$, Eqs.~\eqref{eq:FQ_g_exact_appendix} and \eqref{eq:CFI_g_exact_appendix} reduce to
\begin{equation}
\mathcal F_Q(g)
\simeq
\mathcal F_C(g)
\simeq
\kappa_g^2\frac{4}{\omega_b^2}
\sin^2\left(\frac{\omega_b t}{2}\right)
=
\frac{8me^{2r}}{\hbar\omega\omega_b^2}
\sin^2\left(\frac{\omega_b t}{2}\right).
\label{eq:FQ_CFI_weak_appendix}
\end{equation}
\\ \\

The results above show that $\omega_b$ and $\Omega_g^s\equiv\Omega$ enter the Rabi dynamics
[Eqs.~(B3)--(B5)] as two independent frequencies.  A fit of
the measured $P_1(t)$ to Eq.~(B9) therefore yields both $\omega_b$ and $\Omega$ from the same
data, with no separate calibration of $D$ required. Note that $\omega_b$ only determines $t_{\rm opt}=\pi/\omega_b$; it does not
appear in the relation between $\Omega$ and $g$ below. An imperfect knowledge of $D$ therefore
degrades the choice of interrogation time (i.e., the statistical sensitivity), but does not by
itself bias the inferred value of $g$.
Recall $\Omega=\Omega_g^s=2e^r(mg-F)x_0/\hbar$. Inverting this relation gives
\begin{equation}
g=\frac{F}{m}+\frac{\hbar\,\Omega}{2m x_0 e^r}.
\label{eq:B14}
\end{equation}
This equation shows explicitly that determining $g$ requires knowing three things
independently: the compensation force $F$, the mass $m$ (and hence $x_0=\sqrt{\hbar/2m\omega}$),
and the squeezing parameter $r$.
Rather than relying on $r=\tfrac12\,\mathrm{arctanh}(A_p/\delta)$ alone, the combined factor
$2e^rx_0/\hbar$ in Eq.~(B14) can be calibrated directly. A known auxiliary test force
$\delta F_{\rm test}$ is applied in place of (or added to) gravity — for example via
electrostatic actuation on the levitated particle — and the resulting shift $\delta\Omega$ is
extracted using the fit of measured $P_1(t)$ to Eq.~(B9). Since $\Omega$ is linear in force,
\begin{equation}
\frac{2e^r x_0}{\hbar}=\frac{\delta\Omega}{\delta F_{\rm test}},
\label{eq:B15}
\end{equation}
which fixes the whole scale factor metrologically, without needing $r$ and $x_0$ to be
trusted individually.
Combining Eqs.~(B14)--(B15), the leading-order relative uncertainty in the inferred $g$ is
\begin{equation}
\left(\frac{\delta g}{g}\right)^2\approx
\left(\frac{\delta m}{m}\right)^2+\left(\frac{\delta x_0}{x_0}\right)^2+(\delta r)^2
+\left(\frac{\delta\Omega}{\Omega}\right)^2.
\label{eq:B16}
\end{equation}

\section{Master-equation solution and Fisher information}
\label{app:decoherence_fisher}

We now include mechanical damping in the squeezed-qubit subspace. The zero-temperature damping channel of the original mechanical mode is
\begin{equation}
\dot\rho
=
-\frac{i}{\hbar}[\hat H,\rho]
+
\frac{\gamma_0}{2}\mathcal D_{\hat a,\hat a^\dagger}[\rho].
\end{equation}
Using
\begin{equation}
\hat a=\cosh r\,\hat b-e^{i\theta}\sinh r\,\hat b^\dagger,
\end{equation}
and projecting onto the squeezed-qubit subspace gives
\begin{equation}
\hat a\rightarrow
\hat L_{\rm s}
=
\cosh r\,\hat\sigma_--e^{i\theta}\sinh r\,\hat\sigma_+.
\end{equation}
For the sensing phase $\theta=\pi$, this becomes
\begin{equation}
\hat L_{\rm s}=c\hat\sigma_-+s\hat\sigma_+,
\qquad
c\equiv\cosh r,
\qquad
s\equiv\sinh r.
\end{equation}
The projected master equation is therefore
\begin{equation}
\dot\rho
=
-\frac{i}{2}
[\omega_b\hat\sigma_z+\Omega\hat\sigma_x,\rho]
+
\frac{\gamma_0}{2}
\mathcal D_{\hat L_{\rm s},\hat L_{\rm s}^\dagger}[\rho],
\label{eq:ME_decoh_appendix}
\end{equation}
where $\Omega\equiv\Omega_g^{\rm s}$. Expanding the dissipator gives
\begin{align}
\frac{\gamma_0}{2}\mathcal D_{\hat L_{\rm s},\hat L_{\rm s}^\dagger}[\rho]
\ = \
\frac{\gamma_0}{2}c^2\mathcal D_{\hat\sigma_-,\hat\sigma_+}[\rho]
\ + \
\frac{\gamma_0}{2}s^2\mathcal D_{\hat\sigma_+,\hat\sigma_-}[\rho]
\
+ \ \gamma_0cs
\left(
\hat\sigma_-\rho\hat\sigma_-
+
\hat\sigma_+\rho\hat\sigma_+
\right).
\label{eq:expanded_decoh_appendix}
\end{align}
The first term is relaxation, the second term is excitation induced by the squeezed representation of the bath, and the last line is the phase-sensitive contribution. 
To solve Eq.~\eqref{eq:ME_decoh_appendix}, write the density matrix as
\begin{equation}
\rho(t)=\frac{1}{2}
\left[
\mathds{1}+x(t)\hat\sigma_x+y(t)\hat\sigma_y+z(t)\hat\sigma_z
\right],
\qquad
\bm v(t)=
\begin{pmatrix}
x(t)\\ y(t)\\ z(t)
\end{pmatrix}.
\end{equation}
Using $\hat\sigma_z=|1\rangle_{\rm s}{}_{\rm s}\langle1|-|0\rangle_{\rm s}{}_{\rm s}\langle0|$, the initial state $|0\rangle_{\rm s}$ corresponds to
\begin{equation}
\bm v(0)=
\begin{pmatrix}0\\0\\-1\end{pmatrix}.
\end{equation}
Equation~\eqref{eq:ME_decoh_appendix} is equivalent to
\begin{equation}
\dot{\bm v}(t)=\mathsf A(\Omega)\bm v(t)+\bm b,
\label{eq:bloch_matrix_appendix}
\end{equation}
with
\begin{equation}
\mathsf A(\Omega)=
\begin{pmatrix}
-\Gamma_x & \omega_b & 0\\
-\omega_b & -\Gamma_y & \Omega\\
0 & -\Omega & -\Gamma_z
\end{pmatrix},
\qquad
\bm b=
\begin{pmatrix}0\\0\\-\gamma_0\end{pmatrix},
\label{eq:A_matrix_appendix}
\end{equation}
where
\begin{equation}
\Gamma_x=\frac{\gamma_0}{2}(c-s)^2=\frac{\gamma_0}{2}e^{-2r},
\qquad
\Gamma_y=\frac{\gamma_0}{2}(c+s)^2=\frac{\gamma_0}{2}e^{2r},
\qquad
\Gamma_z=\gamma_0(c^2+s^2)=\gamma_0\cosh(2r).
\end{equation}
The exact solution is
\begin{equation}
\bm v(t)
=
\bm v_{\rm ss}
+
\exp[\mathsf A(\Omega)t]
\left[\bm v(0)-\bm v_{\rm ss}\right],
\label{eq:v_solution_appendix}
\end{equation}
where
\begin{equation}
\bm v_{\rm ss}=-\mathsf A(\Omega)^{-1}\bm b.
\end{equation}
This is the closed-form solution of the master equation; the exponential of the $3\times3$ matrix gives the full damped Rabi dynamics for arbitrary $\Omega$, $\omega_b$, $r$, and $\gamma_0$ within the squeezed-qubit approximation.

For the Fisher information, we need the derivative of $\bm v(t)$ with respect to $\Omega$. Instead of differentiating Eq.~\eqref{eq:v_solution_appendix} explicitly, it is cleaner to propagate the sensitivity vector
\begin{equation}
\bm u(t)
\equiv
\partial_\Omega\bm v(t).
\end{equation}
Differentiating Eq.~\eqref{eq:bloch_matrix_appendix} gives
\begin{equation}
\dot{\bm u}(t)
=
\mathsf A(\Omega)\bm u(t)
+
\mathsf A_\Omega\bm v(t),
\qquad
\bm u(0)=0,
\label{eq:u_equation_appendix}
\end{equation}
where
\begin{equation}
\mathsf A_\Omega
\equiv
\partial_\Omega\mathsf A
=
\begin{pmatrix}
0&0&0\\
0&0&1\\
0&-1&0
\end{pmatrix}.
\end{equation}
Equivalently,
\begin{equation}
\bm u(t)
=
\int_0^t
\exp[\mathsf A(\Omega)(t-\tau)]
\mathsf A_\Omega\bm v(\tau)
\,d\tau .
\label{eq:u_integral_appendix}
\end{equation}
The derivative with respect to $g$ is then
\begin{equation}
\partial_g\bm v(t)
=
\kappa_g\bm u(t),
\qquad
\kappa_g=\frac{2mx_0e^r}{\hbar}.
\end{equation}

For a mixed qubit state with Bloch vector $\bm v$, the QFI with respect to $g$ is
\begin{equation}
\mathcal F_Q^{\rm dec}(g)
=
\left|\partial_g\bm v\right|^2
+
\frac{\left(\bm v\cdot\partial_g\bm v\right)^2}{1-|\bm v|^2}.
\end{equation}
Using $\partial_g\bm v=\kappa_g\bm u$, this becomes
\begin{equation}
\boxed{
\mathcal F_Q^{\rm dec}(g)
=
\kappa_g^2
\left[
\bm u\cdot\bm u
+
\frac{(\bm v\cdot\bm u)^2}{1-\bm v\cdot\bm v}
\right].
}
\label{eq:FQ_decoh_appendix}
\end{equation}
The population measurement in the squeezed-Fock basis, 
$\{|0\rangle_{\rm s},|1\rangle_{\rm s}\}$, is not generally the optimal
measurement in the presence of decoherence. To quantify the maximum
classical information that can be extracted by a qubit projective
measurement, we consider a general readout axis $\bm n$ on the Bloch sphere,
with projectors
\begin{equation}
\Pi_{\pm}^{(\bm n)}
=
\frac{1}{2}\left(\mathbb{I}\pm \bm n\cdot\bm\sigma\right),
\qquad
|\bm n|=1 .
\end{equation}
Writing the density matrix as
\begin{equation}
\rho(t)
=
\frac{1}{2}\left[\mathbb{I}+\bm v(t)\cdot\bm\sigma\right],
\end{equation}
the outcome probabilities are
\begin{equation}
P_{\pm}^{(\bm n)}
=
\frac{1}{2}\left[1\pm \bm n\cdot\bm v(t)\right].
\end{equation}
The corresponding classical Fisher information is therefore
\begin{equation}
\mathcal F_C^{(\bm n)}(g)
=
\frac{
\left[\bm n\cdot\bm u(t)\right]^2
}{
1-\left[\bm n\cdot\bm v(t)\right]^2
}.
\label{eq:CFI_general_axis}
\end{equation}
The conventional population measurement corresponds to $\bm n=\hat{\bm z}$,
namely
\begin{equation}
\mathcal F_C^{(z)}(g)
=
\frac{u_z^2(t)}{1-z^2(t)} .
\end{equation}

The optimized readout is obtained by choosing the measurement axis along the
symmetric-logarithmic-derivative direction. For a mixed qubit state this
direction is
\begin{equation}
\bm n_{\rm opt}
=
\frac{\bm \ell}{|\bm \ell|},
\qquad
\bm \ell
=
\bm u
+
\frac{\bm v\cdot\bm u}{1-|\bm v|^2}\bm v .
\label{eq:nopt}
\end{equation}
We parameterize this axis by the polar and azimuthal angles
\begin{equation}
\bm n_{\rm opt}
=
\left(
\sin\theta_{\rm opt}\cos\phi_{\rm opt},
\sin\theta_{\rm opt}\sin\phi_{\rm opt},
\cos\theta_{\rm opt}
\right),
\label{eq:nopt_angles}
\end{equation}
with
\begin{equation}
\theta_{\rm opt}=\arccos(n_z),
\qquad
\phi_{\rm opt}=\operatorname{atan2}(n_y,n_x).
\end{equation}
The optimized-readout CFI is then
\begin{equation}
\mathcal F_C^{\rm opt}(g)
=
\frac{
\left[\bm n_{\rm opt}\cdot\bm u(t)\right]^2
}{
1-\left[\bm n_{\rm opt}\cdot\bm v(t)\right]^2
}.
\label{eq:CFI_optimized}
\end{equation}
This construction gives the best CFI obtainable
from a two-outcome projective measurement at the chosen operating point.
Physically, this means that decoherence can rotate
or redistribute the information about $g$ among the Bloch-vector components.
A fixed population measurement extracts only the $z$-component of this
information, whereas the optimized readout measures along the direction in
which the state changes most sensitively with respect to $g$.
Experimentally, such a readout can be implemented by applying a final qubit
rotation that maps $\bm n_{\rm opt}\cdot\bm\sigma$ onto $\sigma_z$, followed
by the usual population measurement.
Figure~\ref{fig:nopt_angles_app} displays the locally optimal projective readout used to obtain the optimized CFI in the decoherent calculations.
\begin{figure}[h!]
    \centering
    \includegraphics[width=0.47\linewidth]{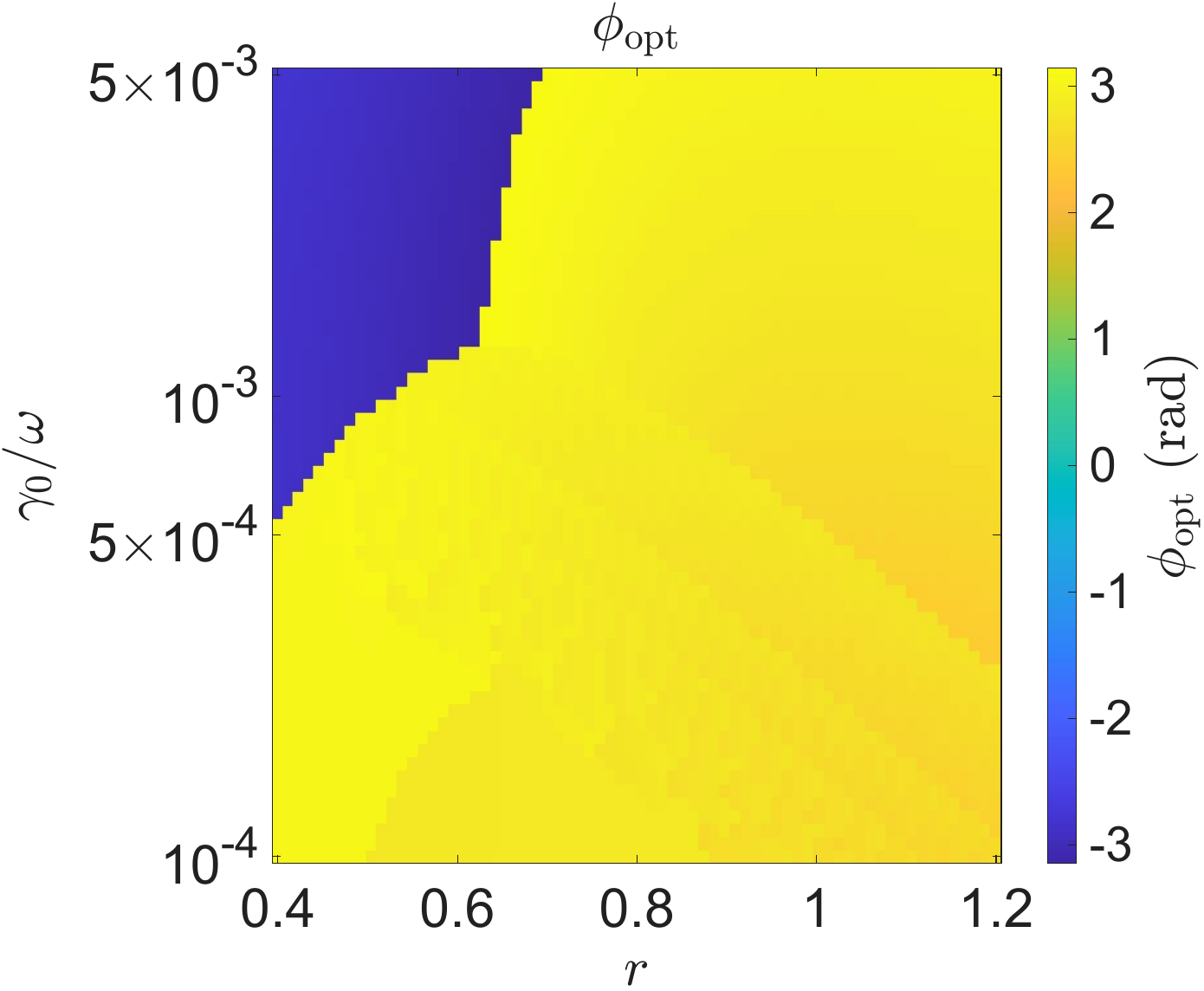}
    \includegraphics[width=0.47\linewidth]{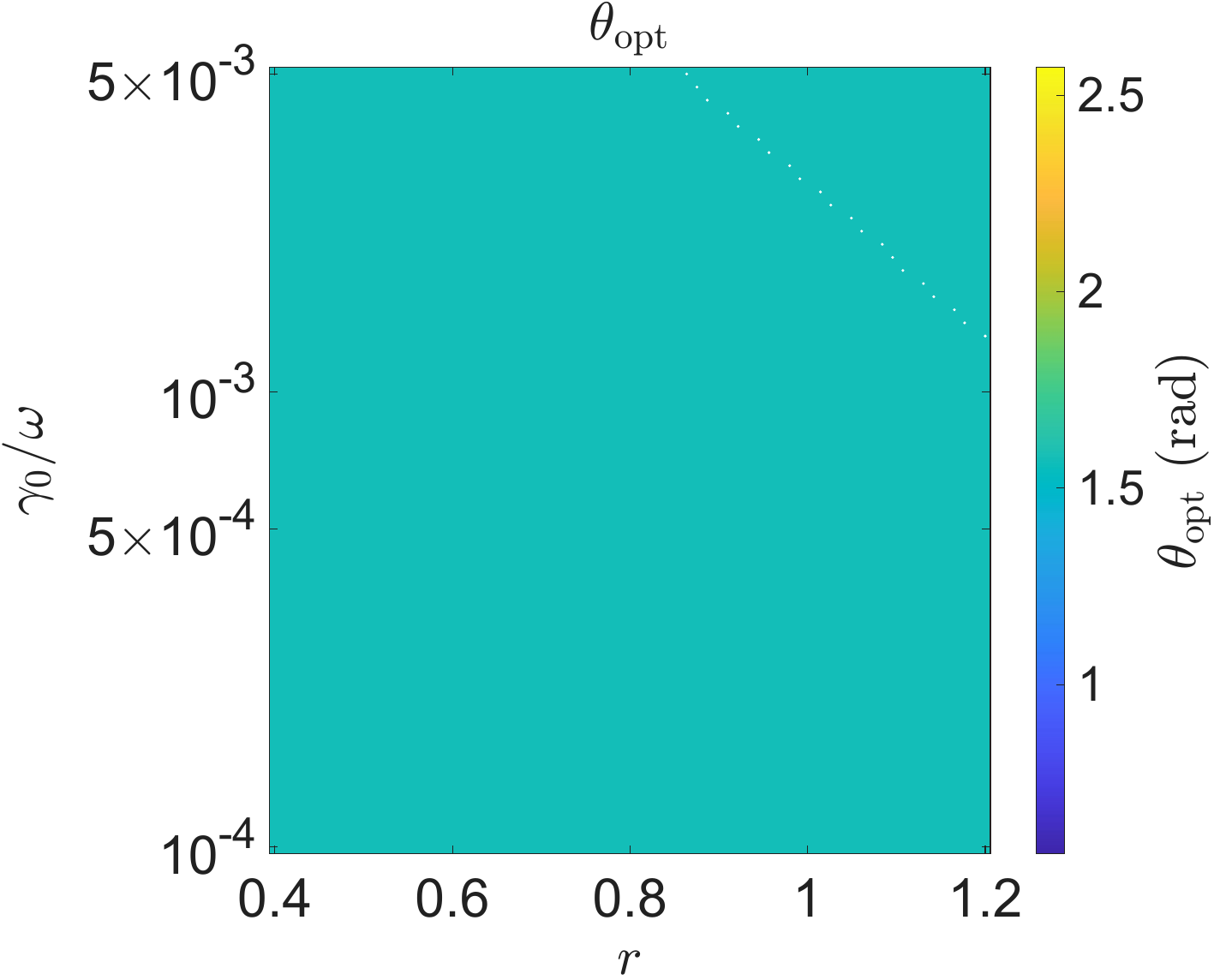}
    \caption{ 
    Corresponding optimal angles $\theta_{\rm opt}$ and $\phi_{\rm opt}$ in ($\gamma_0/\omega, r$) plate, that determine the locally optimal readout direction 
    $\bm n_{\rm opt}{=}(\sin\theta_{\rm opt}\cos\phi_{\rm opt},\sin\theta_{\rm opt}\sin\phi_{\rm opt},\cos\theta_{\rm opt})$ for the decoherent MSF gravimeter. }
    \label{fig:nopt_angles_app}
\end{figure}
\\ \\

Having analyzed the zero-temperature damping model, we now extend to a mechanical bath at finite temperature. In the laboratory frame, the dynamics of the original mechanical mode is governed by
\begin{equation}
    \dot{\rho}
    =
    -\frac{i}{\hbar}\left[\hat{H},\rho\right]
    +
    \frac{\gamma_0}{2}(n_{\mathrm{th}}+1)
    \mathcal{D}_{\hat{a},\hat{a}^{\dagger}}[\rho]
    +
    \frac{\gamma_0}{2}n_{\mathrm{th}}
    \mathcal{D}_{\hat{a}^{\dagger},\hat{a}}[\rho],    
    \label{eq:C_thermal_lab}
\end{equation}
where $ n_{\mathrm{th}} = 1 / {(e^{\left(\hbar\omega/k_{\mathrm{B}}T\right)}-1)}$ is the mean thermal phonon occupation of the mechanical mode with frequency $\omega$, at temperature $T$, with $k_B$ the Boltzmann constant. The two dissipative terms in Eq.~\eqref{eq:C_thermal_lab} describe, respectively, phonon emission and thermal phonon absorption.
Using the Bogoliubov transformation 
and projecting onto the squeezed-qubit subspace for $\theta=\pi$, one obtains the finite-temperature master equation of the MSFQ gravimeter as
\begin{align}
    \dot{\rho}
    =
    -\frac{i}{2}
    \left[
        \omega_b\hat{\sigma}_{z}
        +
        \Omega\hat{\sigma}_{x},
        \rho
    \right]
    +
    \frac{\gamma_0}{2}(n_{\mathrm{th}}+1)
    \mathcal{D}_{\hat{L}_{s},\hat{L}_{s}^{\dagger}}[\rho]
    +
    \frac{\gamma_0}{2}n_{\mathrm{th}}
    \mathcal{D}_{\hat{L}_{s}^{\dagger},\hat{L}_{s}}[\rho],
    \label{eq:C_thermal_master}
\end{align}
where $\Omega\equiv\Omega_g^s$. 
Writing the density matrix in terms of the Bloch vector,
the master equation can again be expressed as
\begin{equation}
    \dot{\mathbf{v}}(t)
    =
    A_T(\Omega)\mathbf{v}(t)
    +
    \mathbf{b}_T,
    \label{eq:C_thermal_bloch}
\end{equation}
with
\begin{equation}
    A_T(\Omega)
    =
    \begin{pmatrix}
        -\Gamma_x^{T} & \omega_b & 0\\
        -\omega_b & -\Gamma_y^{T} & \Omega\\
        0 & -\Omega & -\Gamma_z^{T}
    \end{pmatrix},
    \qquad
    \mathbf{b}_T
    =
    \begin{pmatrix}
        0\\
        0\\
        -\gamma_0
    \end{pmatrix}.
    \label{eq:C_thermal_matrix}
\end{equation}
The finite-temperature anisotropic decoherence rates are
\begin{align}
    \Gamma_x^{T}
    = (2n_{\mathrm{th}}+1)\Gamma_x,   \qquad   
        \Gamma_y^{T}
    = (2n_{\mathrm{th}}+1)\Gamma_y, \qquad
        \Gamma_z^{T}
    = (2n_{\mathrm{th}}+1)\Gamma_z.
    \label{eq:C_thermal_Gz}
\end{align}
Thus, finite temperature multiplies all three anisotropic decay rates by the thermal factor $2n_{\mathrm{th}}+1$. One can follow the previous analysis for $n_{\rm th}{=}0$ to calculate the QFI, the CFI and corresponding optimal measurement. 
The competition between the coherent qubit dynamics and the dominant finite-temperature decay channel is now characterized by
\begin{equation}
    \Xi_T
    \equiv
    \frac{\Gamma_{\mathrm{eff}}^{T}}{\omega_b},
    \qquad
    \Gamma_{\mathrm{eff}}^{T}
    \simeq
    \Gamma_y^{T}
    =
    \frac{\gamma_0}{2}
    (2n_{\mathrm{th}}+1)e^{2r}.
    \label{eq:C_thermal_Xi}
\end{equation}
Again the coherent sensing regime requires $\Xi_T\lesssim1.$
Therefore, finite temperature reduces the accessible squeezing range through the factor $2n_{\mathrm{th}}+1$.
\begin{figure*}[t!]
    \centering
\includegraphics[width=0.24\linewidth]{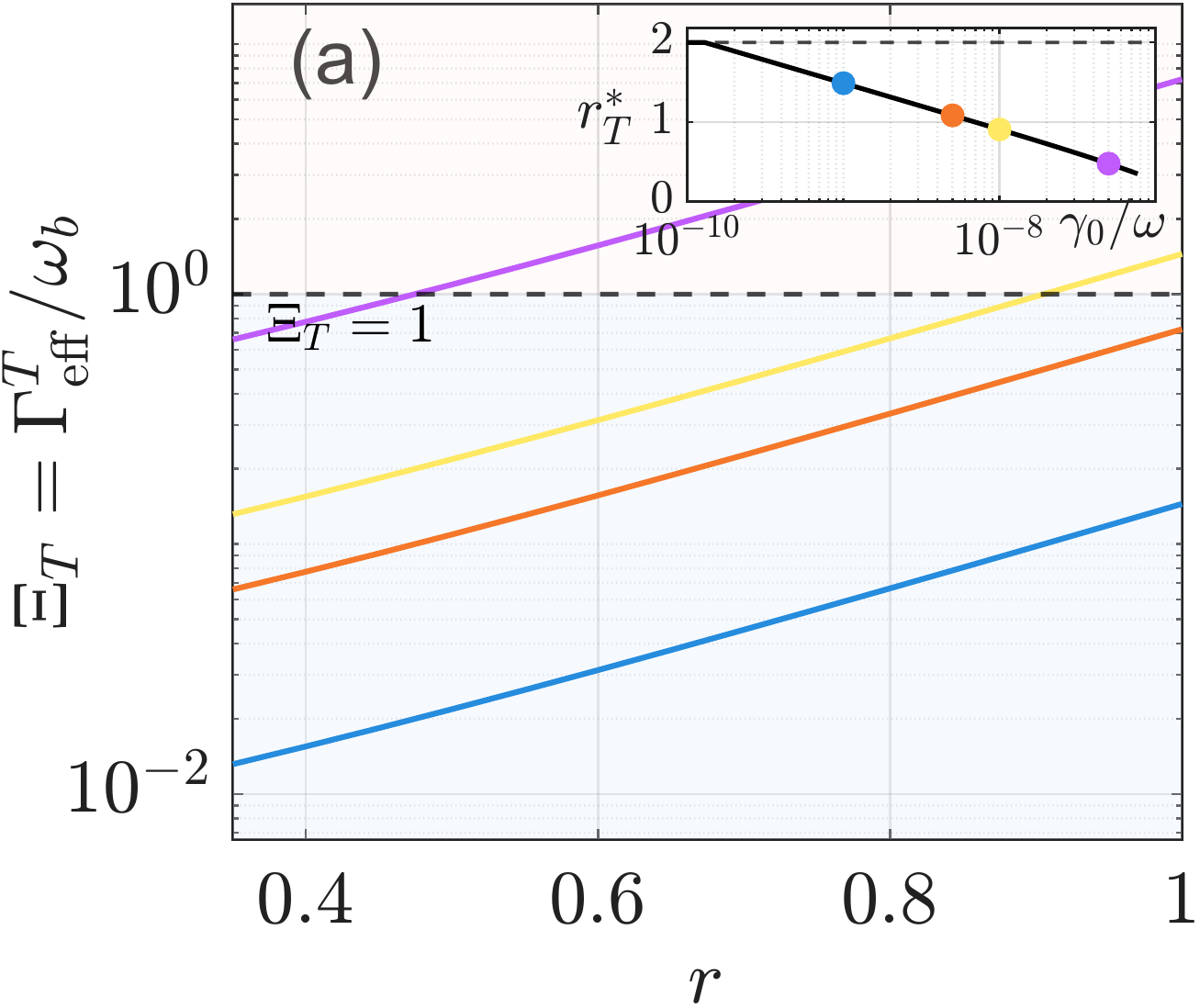}
\includegraphics[width=0.24\linewidth]{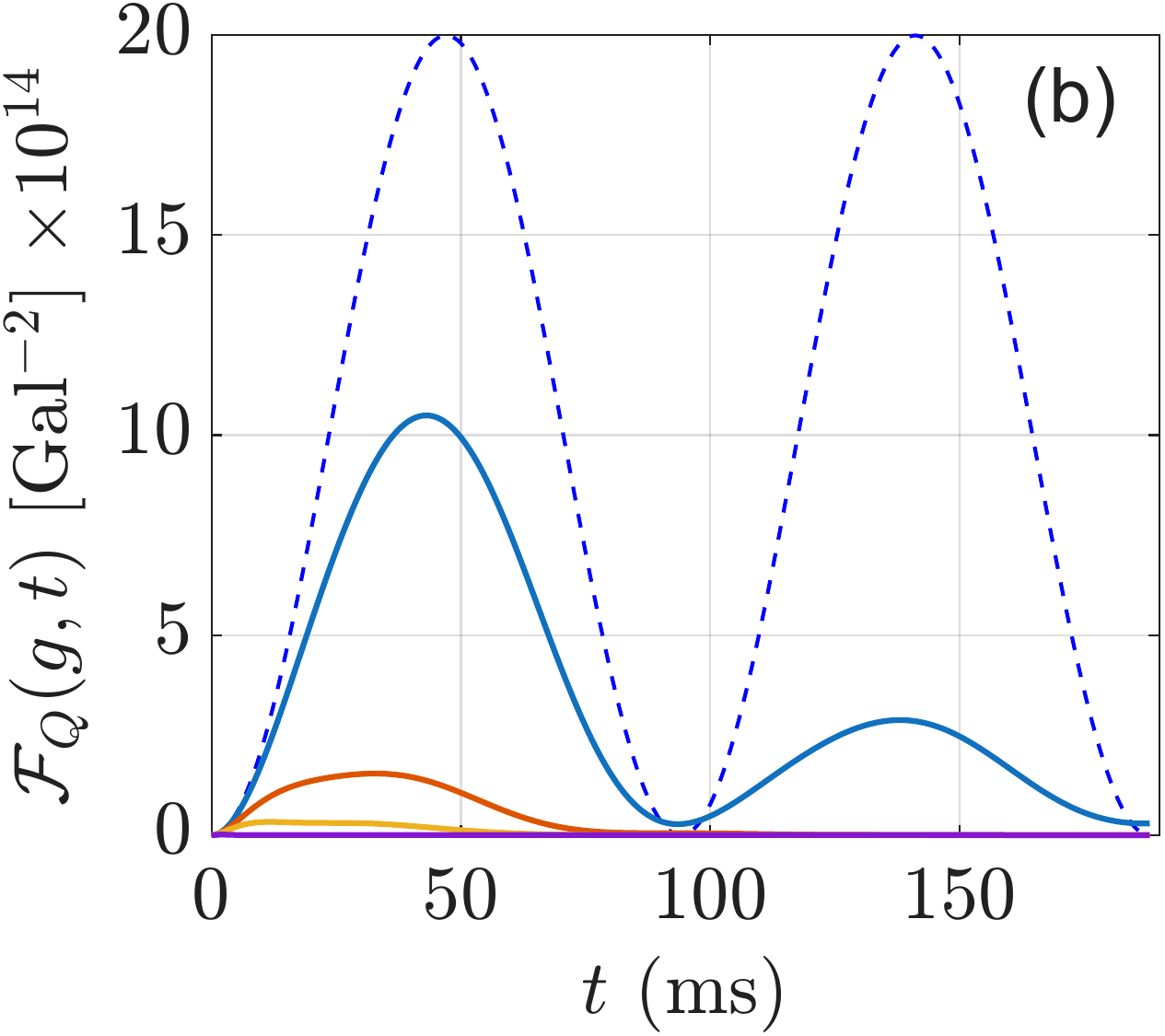}
\includegraphics[width=0.24\linewidth]{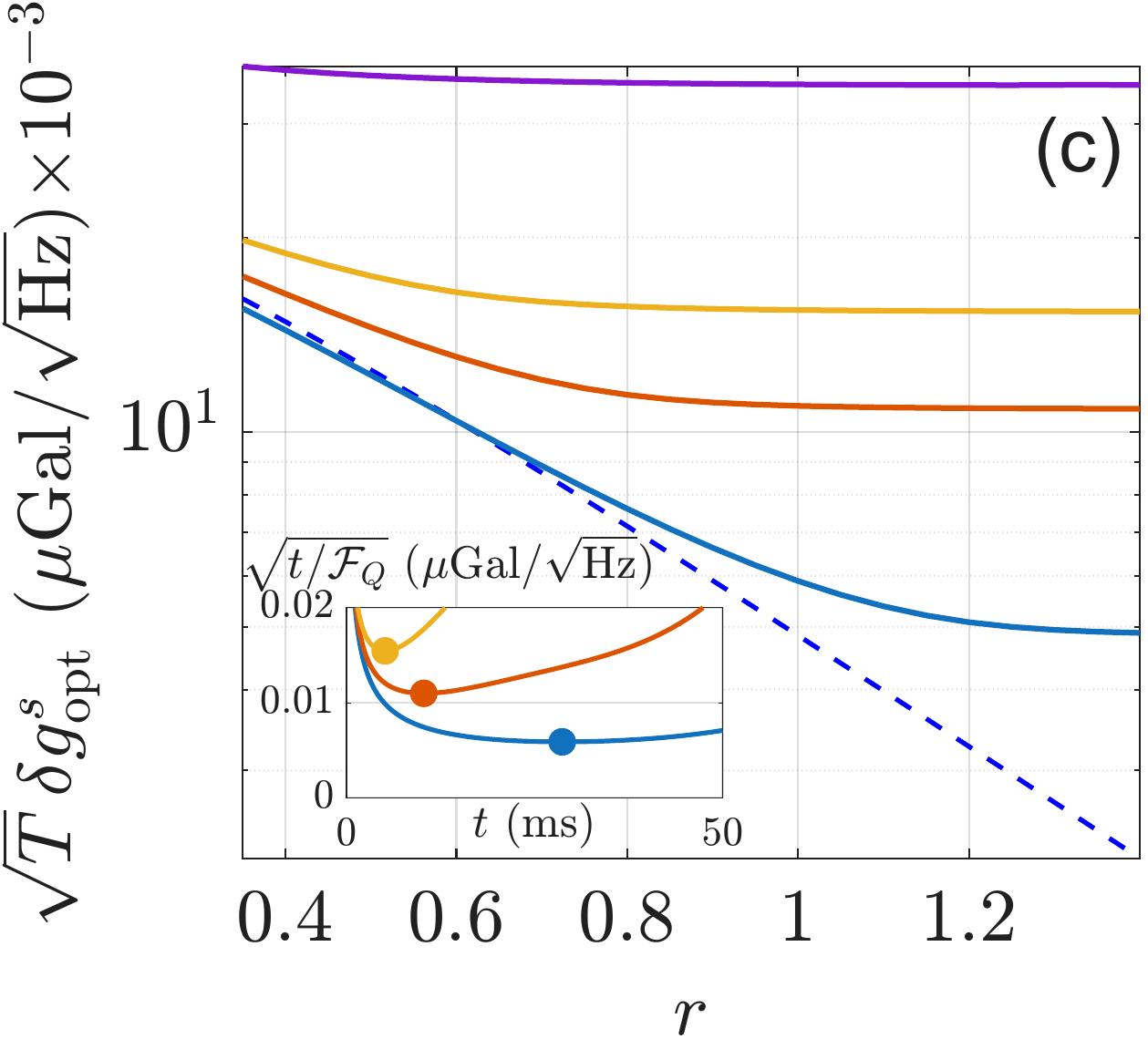}
\includegraphics[width=0.24\linewidth]{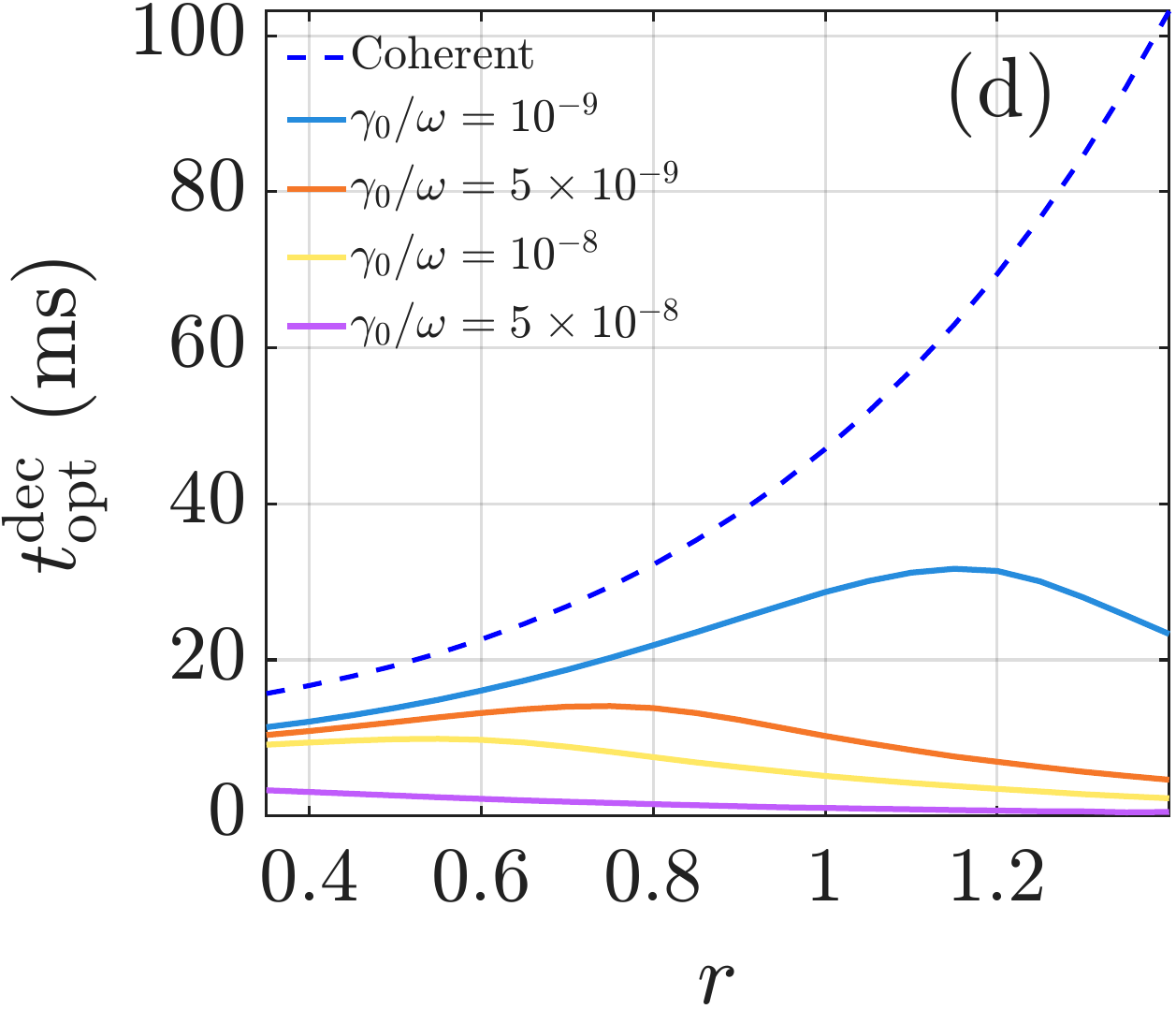}            
\caption{
Finite-temperature performance of the MSFQ gravimeter at a bath temperature of $T=10~\mathrm{mK}$. The intrinsic mechanical damping rates are chosen as $\gamma_0/\omega=10^{-9}$, $5\times10^{-9}$, $10^{-8}$, and $5\times10^{-8}$, corresponding to mechanical quality factors ranging from $Q=10^{9}$ to $2\times10^{7}$. (a) Thermal coherence parameter $\Xi_T=\Gamma_{\rm eff}^{T}/\omega_b$ as a function of the squeezing parameter $r$. The dashed line $\Xi_T=1$ separates the coherent ($\Xi_T<1$) and decoherence-dominated ($\Xi_T>1$) operating regimes. The inset shows the maximum squeezing parameter $r_T^{*}$ satisfying $\Xi_T=1$. (b) QFI as a function of interrogation time. The dashed curve denotes the ideal coherent limit. (c) Optimized time-normalized sensitivity obtained by minimizing $\sqrt{t/F_Q(g,t)}$ with respect to the interrogation time. The dashed curve represents the coherent limit. The inset illustrates the optimization procedure for several intrinsic damping rates, where the minimum of each curve determines the optimal interrogation time. (d) Optimal interrogation time as a function of the squeezing parameter. 
}
\label{fig:A2}
\end{figure*}
Fig~\ref{fig:A2} summarizes the influence of thermal noise on the sensing protocol. 
Fig.~\ref{fig:A2}(a) shows the competition parameter $\Xi_T=\frac{\Gamma_{\rm eff}^{T}}{\omega_b},$
which compares the dominant thermal decoherence rate with the effective qubit frequency. As in the zero-temperature case, the coherent sensing regime is determined by $\Xi_T<1$, whereas $\Xi_T>1$ corresponds to decoherence-dominated dynamics. For all intrinsic damping rates, $\Xi_T$ increases exponentially with the squeezing parameter. The inset shows the maximum squeezing parameter $r_T^*$ satisfying $\Xi_T=1$, which decreases exponentially with increasing damping. Therefore, lower mechanical dissipation enlarges the accessible squeezing window and preserves coherent sensing over a broader parameter regime.
The corresponding QFI is shown in Fig.~\ref{fig:A2}(b). In the absence of decoherence, the QFI exhibits periodic oscillations associated with coherent Rabi dynamics of the MSFQ. Thermal noise suppresses both the peak value of the QFI and its revival amplitude. Nevertheless, for $\gamma_0/\omega=10^{-9}$ the QFI remains appreciable over one interrogation cycle, indicating that coherent information accumulation is still possible even in the presence of a finite-temperature bath.
Fig.~\ref{fig:A2}(c) presents the optimized time-normalized sensitivity obtained by minimizing $\sqrt{t/F_Q(g,t)}$ over the interrogation time. Similar to the zero-temperature case, the coherent limit predicts a monotonic improvement with increasing squeezing because of the exponential enhancement of the gravity-induced coupling. Thermal decoherence eventually limits this improvement by reducing the achievable QFI at large squeezing. For sufficiently weak damping, however, the degradation remains modest over the experimentally relevant squeezing range, and the MSFQ gravimeter continues to outperform conventional mechanical sensing by exploiting the exponential signal amplification provided by the squeezed-Fock qubit. The inset illustrates the optimization procedure by plotting $\sqrt{t/F_Q}$ as a function of interrogation time for several damping rates, where the minimum of each curve determines the optimal operating point.
The corresponding optimal interrogation time is shown in Fig.~\ref{fig:A2}(d). In the ideal coherent limit the optimal interrogation time is simply the half-Rabi period, $t_{\rm opt}^{\rm coh}=\pi/\omega_b$, which increases rapidly with the squeezing parameter because the effective qubit frequency decreases. In contrast, thermal decoherence favors shorter interrogation times in order to reduce information loss. Consequently, the optimal sensing time decreases with increasing intrinsic damping. For weak dissipation ($\gamma_0/\omega=10^{-9}$), the optimal interrogation time remains close to the coherent prediction over a wide squeezing range, whereas stronger damping shifts the optimum toward significantly shorter evolution times. This behavior clearly illustrates the competition between coherent signal accumulation and thermal decoherence that ultimately determines the operating point of the gravimeter.
\\ 
Note that, the damping rates considered in Fig.~\ref{fig:A2}, 
are consistent with the ultrahigh mechanical quality factors $Q={\omega}/{\gamma_0}\simeq2\times10^{7}-10^{9},$ already demonstrated in levitated optomechanical systems~\cite{Gieseler2013NatPhys,Delic2020,Ranjit2016}. 
In particular, quality factors exceeding $10^{8}$ have been experimentally demonstrated for levitated nanoparticles. Therefore, the parameter regime considered here does not rely on unrealistically low mechanical dissipation, but is compatible with the current state of levitated optomechanics.

\end{widetext}
%

\end{document}